\documentclass[fleqn,usenatbib]{mnras}

\usepackage{newtxtext,newtxmath}

\usepackage[T1]{fontenc}
\usepackage{ae,aecompl}

\usepackage{graphicx}	
\usepackage{amsmath}	
\usepackage{multirow}
\usepackage{pdflscape}
\usepackage{array}
\newcolumntype{P}[1]{>{\raggedright\arraybackslash}p{#1}}
\newcolumntype{A}[1]{>{\raggedleft\arraybackslash}p{#1}}

\def\arcsec{\hbox{$^{\prime\prime}$}}
\def\arcmin{\hbox{$^{\prime}$}}

\newcommand{\wsc}{{\tt WSClean}}

\newcommand{\pybdsf}{{\tt pyBDSF}}
\newcommand{\trap}{{\tt TraP}}

\newcommand{\modp}{$V_{\nu}$}
\newcommand{\varp}{$\eta_{\nu}$}
\newcommand{\madp}{$\xi_{\nu,\mathrm{max}}$}
\newcommand{\asec}{$^{\prime\prime}$}
\newcommand{\pcc}{pc cm$^{-3}$}

\newcommand{\gx}{GX\ 339$-$4}
\newcommand{\gxpsr}{PSR\ J1703$-$4851}

\newcommand{\hctwo}{Gaia DR3 5938235492995353728}
\newcommand{\prs}{MKT\ J170121.4$-$493250}
\newcommand{\irgal}{2MASS\ J1701249$-$4932475}

\newcommand{\mtpfrb}{FRB\ 20210405I}

\newcommand*\oline[1]{%
  \vbox{%
    \hrule height 0.5pt
    \kern0.25ex
    \hbox{%
      \kern-0.1em
      \ifmmode#1\else\ensuremath{#1}\fi
      \kern-0.1em
    }
  }
}

\newcommand{\fhs}{\mbox{\ensuremath{.\!\!^{\rm s}}}}

\emergencystretch=\maxdimen
\title[\mtpfrb: a nearby FRB localised with MeerKAT]{\mtpfrb: a nearby Fast Radio Burst localised to sub-arcsecond precision with MeerKAT}

\author[L. N. Driessen et al.]{L. N. Driessen,$^{1,2,3}$\thanks{E-mail: Laura@Driessen.net.au (LND)}
E. D. Barr, $^{4}$
D.\ A.\ H. Buckley, $^{5,6,7}$ 
M. Caleb,$^{3,8}$ 
H. Chen,$^{6,9}$ 
W. Chen,$^{4}$
M. Gromadzki,$^{10}$ 
\newauthor
F. Jankowski,$^{11,2}$ 
R. C. Kraan-Korteweg,$^{6}$ 
M. Kramer,$^{15,2}$
J. Palmerio,$^{12}$ 
K. M. Rajwade,$^{13}$ 
B. W. Stappers,$^{2}$ 
\newauthor
E. Tremou,$^{14}$ 
S. D. Vergani,$^{12}$ 
P. A. Woudt,$^{6}$ 
M. C. Bezuidenhout,$^{15,2}$ 
M. Malenta,$^{2}$ 
V. Morello,$^{2,16}$ 
S. Sanidas,$^{2}$ 
\newauthor
M. P. Surnis,$^{17}$ 
R. P. Fender,$^{18,6}$ 
\\ \\
$^{1}$CSIRO, Space and Astronomy, PO Box 1130, Bentley, WA 6102, Australia\\
$^{2}$Jodrell Bank Centre for Astrophysics, Department of Physics and Astronomy, The University of Manchester, Manchester, M13 9PL, UK\\
$^{3}$Sydney Institute for Astronomy, School of Physics, The University of Sydney, NSW 2006, Australia\\
$^{4}$ Max-Planck-Institut f\"ur Radioastronomie, Auf dem H\"ugel 79, D-53121 Bonn, Germany \\
$^{5}$South African Astronomical Observatory, PO Box 9, Observatory Rd, Observatory 7935, South Africa\\
$^{6}$Department of Astronomy, University of Cape Town, Private Bag X3, Rondebosch 7701, South Africa\\
$^{7}$Department of Physics, University of the Free State, PO Box 339, Bloemfontein 9300, South Africa\\
$^{8}$ASTRO3D: ARC Centre of Excellence for All-sky Astrophysics in 3D, ACT 2601, Australia\\
$^{9}$Research Center for Intelligent Computing Platforms, Zhejiang Laboratory, Hangzhou 311100, China\\
$^{10}$Astronomical Observatory, University of Warsaw, Al. Ujazdowskie 4, 00-478 Warszawa, Poland\\
$^{11}$LPC2E, Universit\'{e} d'Orl\'{e}ans, CNRS, 3A Avenue de la Recherche Scientifique, 45071 Orl\'{e}ans, France\\
$^{12}$GEPI, Observatoire de Paris, Université PSL, CNRS, 5 PlaceJules Janssen, 92190 Meudon, France\\
$^{13}$ASTRON, the Netherlands Institute for Radio Astronomy, Oude Hoogeveensedijk 4, 7991 PD Dwingeloo, The Netherlands\\
$^{14}$National Radio Astronomy Observatory, Socorro, NM, 87801, USA\\
$^{15}$Centre for Space Research, North-West University, Potchefstroom 2531, South Africa \\
$^{16}$SKA Observatory, Jodrell Bank, Lower Withington, Macclesfield, Cheshire, SK11 9FT, UK\\
$^{17}$Department of Physics, Indian Institute of Science Education and Research Bhopal,
Bhopal Bypass Road, Bhauri Bhopal 462 066 Madhya Pradesh, India\\
$^{18}$Department of Physics, Astrophysics, University of Oxford, Denys Wilkinson Building, Keble Road, Oxford OX1 3RH, UK\\
}
\date{Accepted XX. Received YY; in original form ZZ}

\pubyear{2022}

\begin{document}
\label{firstpage}
\pagerange{\pageref{firstpage}--\pageref{lastpage}}
\maketitle

\begin{abstract}
We present the first sub-arcsecond localised Fast Radio Burst (FRB) detected using MeerKAT. {\mtpfrb} was detected in the incoherent beam using the MeerTRAP pipeline on 2021 April 05 with a signal to noise ratio of 140.8 and a dispersion measure of 565.17 pc cm$^{-3}$. It was detected while MeerTRAP was observing commensally with the ThunderKAT large survey project, and was sufficiently bright that we could use the ThunderKAT 8s images to localise the FRB. 
Two different models of the dispersion measure in the Milky Way and halo suggest that the source is either right at the edge of the Galaxy, or outside.
{This highlights the uncertainty in the Milky Way dispersion measure models, particularly in the Galactic Plane, and the uncertainty of Milky Way halo models. Further investigation and modelling of these uncertainties will be facilitated by future detections and localisations of nearby FRBs.}
We use the combined localisation, dispersion measure, scattering, specific luminosity {and chance coincidence probability information} to find that the origin is most likely extra-galactic and identify the likely host galaxy of the FRB: \irgal. Using SALT spectroscopy and archival observations of the field, we find that the host is a disk/spiral galaxy at a redshift of $z=0.066$.
\end{abstract}




\section{Introduction}
\label{sec: introduction}

Fast radio bursts (FRBs) are short, bright, coherent radio transients with durations on the order of milliseconds \citep[see e.g.][for recent reviews]{2019ARA&A..57..417C,2019A&ARv..27....4P,2021arXiv210710113P,2021Univ....7..453C,2021A&G....62.1.29C}. Most FRBs are observed as one-off events, though some are observed to repeat. Due to their transient nature and short duration, localising one-off events to less than a few arcseconds remains a challenge.

{The first FRB to be localised, FRB~20121102A, was a repeater. This repeating nature allowed it to ultimately be localised to 0\farcs1 \citep{2017Natur.541...58C,2017ApJ...834L...8M} and associated it with a faint, continuum radio source which exhibits flux variability of tens of percent on day timescales. \citet{2017ApJ...834L...7T} imaged the position of the FRB with the Gemini North telescope, revealing a dwarf galaxy at a redshift of z=0.19273(8). This galaxy has a low metallicity and shows no signs of AGN activity. Very long baseline interferometry (VLBI) follow-up observations with milli-arcsecond resolution revealed a project linear separation of $\lesssim 40$~pc between the source of the FRBs and the persistent continuum radio source with the latter having a projected size of $\lesssim 0.7$~pc \citep{2017ApJ...834L...8M}. All together, the observations suggest that the source of the FRBs could be associated with a low-luminosity active galactic nucleus or a young neutron star powering a supernova remnant \citep{2017Natur.541...58C,2017ApJ...834L...8M}. The case of FRB~20121102A demonstrates how important localisation is for identifying host galaxies and unveiling local environments, ultimately shedding light on the physical processes leading to the burst production. }

{With the advent of intereferometeric searches, FRBs are now almost routinely being localised upon discovery without having to rely on repeaters for precise localisation. There were $\sim$800 published FRBs in the Supernova Working Group Transient Name Server\footnote{\href{https://www.wis-tns.org/}{https://www.wis-tns.org/}} (TNS) at the time of writing, of which {57} are known repeaters. }
Presently, {{27}
FRBs have been conclusively localised to a host galaxy}. 
{Nineteen} of these FRBs 
\citep{2022AJ....163...69B,Bannister565,2019Sci...366..231P,2020Natur.581..391M,2020ApJ...903..152H,2020ApJ...901L..20B,2022MNRAS.516.4862J,2022arXiv221004680R,2023ApJ...948...67B,2023arXiv230205465G} were detected and localised using the Australian Square Kilometre Array Pathfinder
\citep[ASKAP;][]{2021PASA...38....9H}, four 
\citep{2017Natur.541...58C, 2020ApJ...899..161L,2022Natur.606..873N,2020Natur.586..693L}
have been localised using the Karl G. Jansky Very Large Array \citep[VLA;][]{perley2011}, two 
\citep{2021arXiv210301295B, 2022Natur.602..585K, 2021ATel14603....1M, 2022ApJ...927L...3N} 
have been localised using Very Long Basline Interferometry (VLBI) {both} using the European VLBI network 
\citep[EVN;][]{2020Natur.577..190M}, one has been localised using the Deep Synoptic Array 10-dish (DSA-10) prototype 
\citep{2019Natur.572..352R}, {and one \citep{2023ATel16191....1R} has been localised with the DSA-110 prototype}. {Seven} of the {27} localised FRBs are repeaters.
Upon investigating the global properties of a sample of FRB host galaxies, \citet{2022AJ....163...69B} {and \citet{2023arXiv230205465G}} found no significant difference between the host galaxies of repeating and non-repeating FRBs.

Localisation reveals important information about FRBs and their hosts, and gives us clues about their progenitors {and immediate environments.} 
Some progenitor models for FRB~20121102A include a young neutron star in a compact supernova remnant or a neutron star near the accretion torus of a black hole \citep[][]{2017Natur.541...58C,2018Natur.553..182M,2018ApJ...854L..21Z}.
The localisation of the repeating FRB~20200120E \citep{2022Natur.602..585K} to a globular cluster in the M81 galactic system
{requires an unusual} magnetar progenitor model for FRBs as globular clusters consist of older stars, not the young population expected for magnetars.
If we can determine the position of an FRB {progenitor}, we can also determine if the FRB passed through the halo of an intervening galaxy. \citet{2019Sci...366..231P} used {one such event} FRB to constrain the halo gas density, magnetisation and turbulence of the intervening galaxy. \citet{2020Natur.581..391M} demonstrated that a large sample of arcsecond-localised FRBs can be used to directly measure the baryons in the Universe.

MeerKAT \citep{2018ApJ...856..180C} is an interferometer in the Karoo region of South Africa. It consists of 64$\times$13.5\ m dishes and has baselines up to 8\ km, resulting in a field of view (FoV) of $\sim$1\ square\ degree and resolution of better than $\sim5\arcsec$ at 1400\ MHz. MeerKAT currently has two receivers installed, the L-band receiver (856-1712\ MHz) and the UHF receiver (580-1015\ MHz) and can observe in modes with 1024, 4096, or 32,000 channels.

The Meer(more) TRAnsients and Pulsars \citep[MeerTRAP;][]{2018IAUS..337..406S,2022MNRAS.512.1483B,2022ASPC..532..273J,2022MNRAS.514.1961R} project at the MeerKAT telescope undertakes fully commensal, high time resolution radio transient searches simultaneously with all of the ongoing MeerKAT Large Survey Projects (LSPs).

ThunderKAT \citep{2017arXiv171104132F} is the MeerKAT LSP searching for, and investigating, radio transient and variable sources in the image-plane.
ThunderKAT observes and monitors known variable radio sources such as X-ray binary systems 
\citep[e.g.][]{2019ApJ...883..198R,2020NatAs...4..697B,2020MNRAS.491L..29W,2022MNRAS.516.2641V},
follows up sources that were initially detected at other wavelengths \citep[][]{2020MNRAS.496.3326R,2020MNRAS.496.2542H,2022MNRAS.510.1083D}, and performs untargeted searches for transient and variable sources in the FoV of MeerKAT image-plane observations \citep[][]{FlareyBoi,2022arXiv220309806D,2022MNRAS.513.3482A,2022arXiv220316918R}.

In this paper we present the detection and localisation {of \mtpfrb\ }the first sub-arcsecond localisation of an FRB using MeerKAT. In Section\ \ref{sec: radio observations} we describe the MeerTRAP and ThunderKAT radio observations. In Section\ \ref{sec: detection of mtpfrb} we present the detection and properties of \mtpfrb. In Sections\ \ref{sec: discussion} 
and \ref{sec: conclusions} we {summarise} and conclude.


\section{Radio observations}
\label{sec: radio observations}


\subsection{MeerTRAP pipeline}
\label{sec: meertrap pipeline}

For the observations presented in this paper, MeerKAT was observing at an L-band centre frequency of 1284 MHz with a usable bandwidth of $\sim770$~MHz.
Single pulses are searched for simultaneously in incoherent and coherent modes using the MeerTRAP backend.
In the coherent mode, the voltages from the inner 40--44 dishes of the $\sim1$-km core of the array are coherently combined to {764} coherent beams (CBs) on sky with a total FoV of $\sim0.4$ deg$^2$. A CB typically {has a FWHM of} $\sim43\arcsec$ but can vary in size depending on the elevation of the telescope at the time of observation. In the incoherent mode the intensities of all the available 64 MeerKAT dishes are summed to create a less sensitive incoherent beam (IB) but with a much wider FoV of $\sim1.3$ deg$^2$. The CBs typically tile the IB with beams touching where the sensitivity is 25\% of the peak CB sensitivity.

The output data stream of the F-engine are captured, delay corrected, phased and channelised before being sent over the network to the beamforming User Supplied Equipment (FBFUSE) that was designed and developed at the Max Planck Institute for Radio Astronomy in Bonn.
Typically, for L-band observations FBFUSE combines the data into 764 total-intensity tied-array beams.
The data are then captured at 306.24 $\upmu$s time resolution by the Transient User Supplied Equipment (TUSE), a real-time transient detection backend instrument developed by the MeerTRAP team at the University of Manchester.
The GPU-based single pulse search pipeline AstroAccelerate\footnote{\href{https://github.com/AstroAccelerateOrg/astro-accelerate}{https://github.com/AstroAccelerateOrg/astro-accelerate}} \citep{2012ASPC..461...33A,2020ApJS..247...56A} is used to search for bursts in real-time after incoherently de-dispersing the data in the DM range 0--5118.4~pc cm$^{-3}$ \citep[see][for more details]{csa+20}.


\subsection{ThunderKAT imaging}
\label{sec: thunderkat imaging}

The ThunderKAT observations presented here are targeted observations of the low-mass X-ray binary \gx. ThunderKAT has been observing this source on a weekly cadence since September 2018 and will continue to do so until August 2023 \citep{2020MNRAS.493L.132T}. The high cadence and length of the monitoring campaign make this an excellent field for {performing comemensal searches for} variable and transient sources \citep[][]{FlareyBoi,2022arXiv220309806D}.

The ThunderKAT \gx\  observations are taken using the MeerKAT L-band receivers in full-polarisation mode. The shortest integration time available for most imaging observations is 8\ s. The L-band receivers have a bandwidth of 856\ MHz with a central frequency of 1284\ MHz. The bandpass and flux calibrator, {PKS\ J1934$-$638}, is observed for 5\ min at the start of the observing block and the phase calibrator, {PMN~J$1726-5529$}, is observed for 2\ min before and after the target scan. {The target is typically observed for 10\ min.}

The imaging data were processed using standard methods including flagging with \texttt{AOFlagger}\footnote{\href{https://sourceforge.net/projects/aoflagger/}{https://sourceforge.net/projects/aoflagger/}}
\citep{2010offringa,2012aoflagger} and calibration, including phase correction, antenna delays and band-pass corrections, using the Common Astronomy Software Application\footnote{\href{https://casa.nrao.edu/}{https://casa.nrao.edu/}} \citep[CASA;][]{2007ASPC..376..127M}. Further details on the processing methods can be found in \citet{FlareyBoi} and \citet{2020MNRAS.493L.132T}.

In order {to image as well as possible} over short time intervals, we produced an {image mask} using a deep, combined MeerKAT image of the \gx\  field made by jointly imaging the visibilities with \texttt{DDFacet} \citep{2018tasse} from 8 epochs (a commissioning observation from April 2018 and the weekly epochs from September and October 2018) for a total integration time of 3.63\ hours. 
The data were then imaged with \wsc\ \citep{offringa-wsclean-2014} utilising the {image mask}. Images of the single epoch, full integration time as well as the shortest integration time (8\ s), were made. 
We included multi-scale cleaning, a Briggs robust weight of $-0.7$ \citep[][]{1995AAS...18711202b} and three $w$-projection plane layers. To maximise the signal-to-noise ratio (S/N) we performed a multi-frequency synthesis (MFS) clean using 8 frequency channels and a 4$^{\mathrm{th}}$ order spectral polynomial fit. We also produced subband images by removing the spectral fit and MFS imaging parameters, resulting in 8 subband images with central frequencies: 909, 1016, 1123, 1230, 1337, 1444, 1551, 1658\ MHz. We then primary beam corrected the sub-band images by {dividing} them by primary beam models. 

Searching for variables and transients in ThunderKAT images was performed using the the \texttt{LOFAR Transients Pipeline} \citep[\trap, Release 4.0;][]{Swinbank2015}.
The \trap\ is software designed with LOFAR\footnote{The transients key project: \href{https://transientskp.org/}{https://transientskp.org/}} specifically in mind. It extracts light curves for radio sources from a time series of fits images. 
We used the default \trap\  parameters\footnote{The default settings for the pipeline configuration and job configuration files can be found in the \href{https://tkp.readthedocs.io/en/latest/userref/config/}{\trap\  documentation}.} with some small adjustments.
{We forced the \trap\ to search for sources consistent with the Gaussian synthesised beam shape and required that sources be separated by three beamwidths to be considered unique. }
In other work searching for variable sources in this field \citep[see][]{2022arXiv220309806D} we manually identified resolved sources and artefacts. We removed these sources in our analysis\footnote{See \href{https://github.com/AstroLaura/MeerKAT\_LightCurve\_Analysis}{https://github.com/AstroLaura/MeerKAT\_LightCurve\_Analysis} for the code for extracting the source information from \trap.}.

We searched for interesting sources in the \trap\ light curves using three variability parameters: \varp, \modp, and \madp.
The \varp\ parameter is based on the reduced $\chi^2$ statistic, {where the model light curve assumes that the flux density is constant}:
\begin{eqnarray}
    \eta_{\nu} & = & \frac{1}{N-1}\sum^{N}_{i=1}\frac{\left(I_{\nu,i}-\oline{I_{\nu}}\right)^{2}}{\sigma^{2}_{\nu,i}} \nonumber \\
     & = & \frac{N}{N-1}\left(\oline{wI^{2}} - \frac{\oline{wI}^{2}}{\oline{w}} \right)
    \label{eq: VAR chi2 param}
\end{eqnarray}
where N is the number of measurements, $I_{\nu,i}$ is the flux density {for epoch i}, \oline{I_{\nu}} is the mean {flux density} and $w$ is the weight {($w=1/{\sigma^{2}_{\nu,i}}^{2}$ where ${\sigma^{2}_{\nu,i}}$ are the $1\sigma$ uncertainties)}. We therefore expect a constant source to have a \varp\ close to one.
The \modp\ parameter is the standard deviation of the light curve divided by the mean of the light curve and as such measures the spread of the flux density measurements. 
The \madp\ parameter is based on the median and the median absolute deviation (MAD) of the light curve where the MAD is given by:
\begin{eqnarray}
    \mathrm{MAD} = \mathrm{median}\left(\left| I_{\nu,i} - \widetilde{I_{\nu}} \right| \right)
    \label{eq: MAD definition}
\end{eqnarray}
where $\widetilde{I_{\nu}}$ is the median flux density.
{The $\xi$ value is then calculated using:}
\begin{eqnarray}
    \xi_{i} & = & \frac{I_{\nu,i} - \widetilde{I_{\nu}}}{\mathrm{median}\left(\left| I_{\nu,i} - \widetilde{I_{\nu}} \right| \right)}
\end{eqnarray}
and the \madp\  value is the resulting maximum value of the set $\xi_{i}$. A high \madp\ indicates a short, bright outburst and is useful for detecting transients.




\subsubsection{Absolute astrometry}
\label{sec: absolute astrometry}

We corrected the absolute astrometry of the radio sources in the \gx\  field using the method described in \citet{2022arXiv220309806D}. We used the Python Blob Detector and Source Finder\footnote{\href{https://www.astron.nl/citt/pybdsf/}{https://www.astron.nl/citt/pybdsf/}} (\pybdsf) to determine the positions of sources in an image, which we used to determine and correct the accuracy of our absolute astrometry. We used Australian Telescope Compact Array (ATCA) Parkes-MIT-NRAO (PMN) sources \citep[ATPMN;][]{2012MNRAS.422.1527M} within the MeerKAT FoV to correct the astrometry. The ATPMN survey has a median absolute astrometric uncertainty of 0\farcs4 in both Right Ascension (RA) and Declination (Dec) {when compared to the Long Baseline Array (LBA) Calibrator Survey \citep[LCS1;][]{2011MNRAS.414.2528P} and the International Celestial Reference Frame \citep[ICRF;][]{1998AJ....116..516M}. Both LCS1 and ICRF use the International Celestial Reference System (ICRS).}

{To determine our astrometric accuracy, we solved for a transformation matrix, $A$ to shift and rotate the MeerKAT sources to match the positions of the ATPMN sources\footnote{The code for performing the astrometric corrections can be found on GitHub: \href{https://github.com/AstroLaura/MeerKAT\_Source\_Matching}{https://github.com/AstroLaura/MeerKAT\_Source\_Matching}}. 
If $X$ represents the uncorrected coordinates and $X^{'}$ represents the corrected coordinates, then we are solving for $A$:
\begin{align}
    X^{'} & = XA \\
    \begin{bmatrix}
    RA_1^{'} & DEC_1^{'} & 1 \\
    RA_2^{'} & DEC_2^{'} & 1 \\
    RA_N^{'} & DEC_N^{'} & 1 \\
    \end{bmatrix} & = \begin{bmatrix}
    RA_1 & DEC_1 & 1 \\
    RA_2 & DEC_2 & 1 \\
    RA_N & DEC_N & 1 \\
    \end{bmatrix} \begin{bmatrix}
    a_1 & a_2 & 1 \\
    b_1 & b_2 & 1 \\
    c_1 & c_2 & 1 \\
    \end{bmatrix}
\end{align}
where $A$ is an affine transformation matrix that includes scale, shear and translation.}

We then applied $A$ to the coordinates of all sources in the field.
In the case of the 8\ s integration images, we determined the astrometric correction for the full integration (usually 10\ min) image and applied that correction to the 8\ s slices.


\section{\mtpfrb}
\label{sec: detection of mtpfrb}

\begin{figure}
\includegraphics[width=\columnwidth]{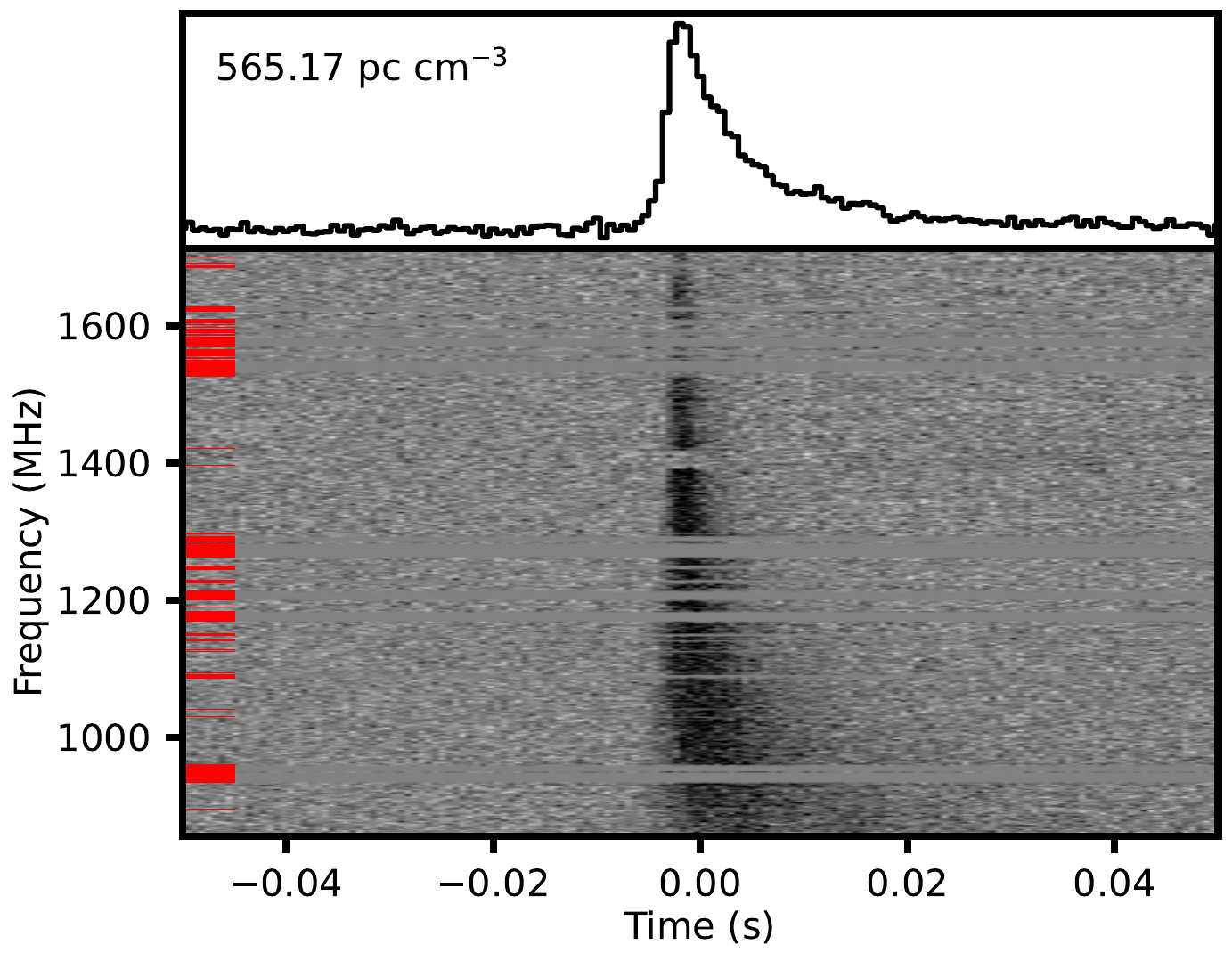}
\caption{Dynamic spectrum of \mtpfrb. The lower panel shows the time and frequency.
The top panel shows the frequency averaged burst profile. {The profile has been de-dispersed to the scattering-corrected DM of 565.17 pc cm$^{-3}$. The S/N-maximising DM is 566.43 pc cm$^{-3}$.
The red lines indicate frequency channels that have been masked due to RFI.
The dropout of the FRB signal around 1.4~GHz that can be seen in both the dynamic and the total intensity spectrum in Figure~\ref{fig:scattering} is instrumental in nature. The time samples in that $\sim$30~MHz frequency range are delayed by about 1.25~s with respect to the rest of the data. This is a known issue that used to happen in exceptional cases when one or more beamformer nodes dropped out of sync with the rest of the cluster, thereby delaying their signal. The frequency width of the shifted signal of $2 \times 856~\text{MHz} / 64 = 26.75~\text{MHz}$ indicates that two logical nodes (one physical server) were affected.}
}
\label{fig: MTP19 waterfall}
\end{figure}

\mtpfrb\ was detected while MeerTRAP was observing commensally with ThunderKAT during an observation of \gx\ on 2021 April 05 at 04:14:40 UTC. The pulse, shown in Figure\ \ref{fig: MTP19 waterfall}, was extremely bright, and was detected by the MeerTRAP real-time transient pipeline in the IB as well as several hundred CBs, implying a nearby origin. The pulse was brightest in the IB, and we obtain an optimized S/N of 140.8 corresponding {to a S/N-maximising DM of 566.43 pc cm$^{-3}$ using the \textsc{mtcutils}\footnote{\href{https://bitbucket.org/vmorello/mtcutils/}{https://bitbucket.org/vmorello/mtcutils/}} package or a scattering-corrected DM of 565.17\ \pcc}. {The scattering-corrected DM is the DM which gives the minimum pulse width when scattering, intrinsic width and dispersion measure smearing has been accounted for.} The expected Galactic DM contribution is $\sim350$ pc cm$^{-3}$ using the \textsc{ymw16} Galactic free-electron model \citep{YMW16} or $\sim500$ pc cm$^{-3}$ using the \textsc{ne2001} Galactic free-electron model \citep{2002Cordes}. 
We assume a conservative halo DM contribution of {$144\pm60$\ \pcc\ }(with a 40\ per\ cent uncertainty) as the source is close to the Galactic plane \citep{2020Yamasaki}. Here we will present the properties of \mtpfrb, with a summary shown in Table\ \ref{tab: mtp19 properties}.

\begin{table}
\caption{Properties of \mtpfrb.}
\label{tab: mtp19 properties}
\begin{tabular}{lcc}
\hline
FRB	                    &                                   & \\
Parameter               & Unit & \\
\hline
\multicolumn{3}{c}{Measured}\\
MJD$_\text{topo}^\text{a}$  &                               & 59309.1768572970 \\
UTC$_\text{topo}^\text{a}$  &                               & 2021-04-05 04:14:40.470 \\
Beam                    &                                   & IB\\
RA                      & (hms)                             & 17$^{\mathrm{h}}$01${\mathrm{m}}$21\fhs5 $\pm$0\farcs4 \\
Dec                     & (dms)                             & $-49^{\circ}$32\arcmin42\farcs8 $\pm$0\farcs5\\
l                       & (deg)                             & 338.1920\\
b                       & (deg)                             & -4.5969\\
S/N-maximising DM       & ($\text{pc} \: \text{cm}^{-3}$)   & 566.43\\
Scattering-corrected DM & ($\text{pc} \: \text{cm}^{-3}$)   & 565.17\\
$\text{S/N}$   &                                   & 140.8\\
W$_\text{50p}^\text{a}$ & (ms)                              & $9.79\pm0.23$\\
W$_\text{10p}^\text{a}$ & (ms)                              & $23.80\pm0.53$\\
$\tau_s^\text{a}$       & (ms)                              & $8.67\pm0.28$\\
W$_\text{eq}^\text{a}$  & (ms)                              & $11.81\pm0.23$\\
Project physical offset & (kpc)                             & 10.65 \\
\hline
\multicolumn{3}{c}{Instrumental}\\
$t_\text{dm}^\text{b}$  & (ms)                              & 6.3 \\ 
$N_\text{chan}$         &                                   & 1024 \\
$N_\text{ant,ib}$       &                                   & 60 \\
LSP                     &                                   & ThunderKAT \\
\hline
\multicolumn{3}{c}{Inferred}\\
$S_\text{peak}$     & (Jy)                                 & 15.9\\
$F$                 & (Jy ms)                               & 120.8\\
DM$_{\text{mw,ne2001}}$  & ($\text{pc} \: \text{cm}^{-3}$)  & 516.1\\
DM$_{\text{mw,ymw16}}$  & ($\text{pc} \: \text{cm}^{-3}$)   & 348.7\\
DM$_{\text{halo}}$  & ($\text{pc} \: \text{cm}^{-3}$)       & 144\\
DM$_\text{xg,ne2001}$      & ($\text{pc} \: \text{cm}^{-3}$)       & 50.3\\
DM$_\text{xg,ymw16}$      & ($\text{pc} \: \text{cm}^{-3}$)       & 217.7\\
$z$                 &                                       & 0.066\\
\hline
\multicolumn{3}{l}{$^\text{a}$ Measured at 1016.5~MHz.}\\
\multicolumn{3}{l}{$^\text{b}$ Intra-channel dispersive smearing in the lowest frequency channel.}\\
\end{tabular}
\end{table}



\subsection{Localisation}
\label{sec: localisation of mtpfrb}

The high detection S/N of \mtpfrb\  meant that we could utilise the ThunderKAT imaging observations to localise it. The imaging observation had an observation time of 10\ min, with a minimum integration time of 8\ s. {The pulse has a dispersion delay of $\sim1.4$\ s, which is fully contained within one 8\ s image.} As described in Section\ \ref{sec: thunderkat imaging}, we made a full time integration MFS image and $74\times8$\ s image slices using \wsc.

\begin{figure*}
\includegraphics[width=\textwidth]{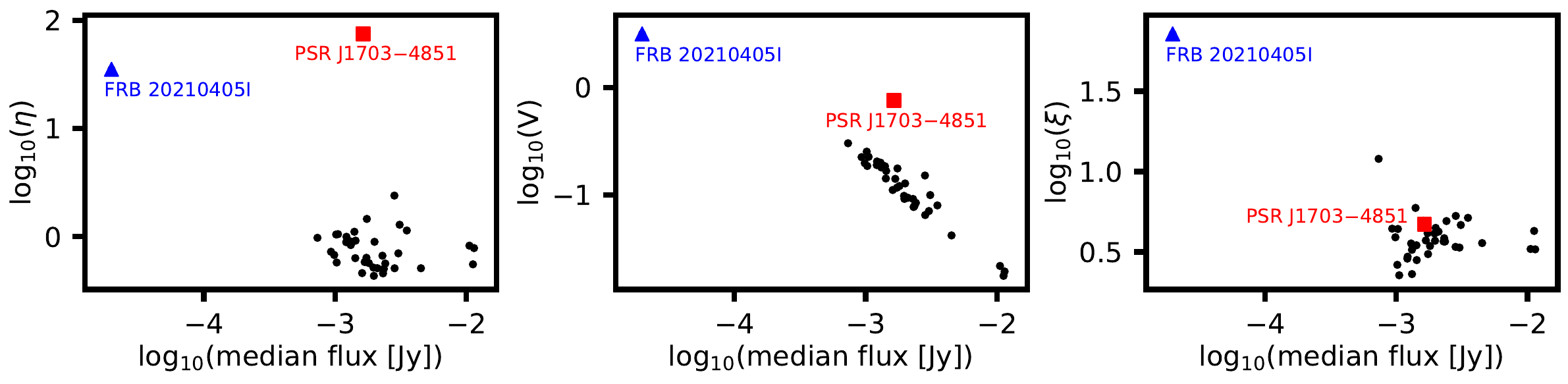}
\caption{Variability parameters (described in Section\ \ref{sec: thunderkat imaging}) for all sources in the 8\ s images from 2021 April 05. \gxpsr\ (labelled) is an outlier. The other, lower flux density outlier is \mtpfrb.
}
\label{fig: MTP19 var params}
\end{figure*}

\begin{figure*}
\includegraphics[width=\textwidth]{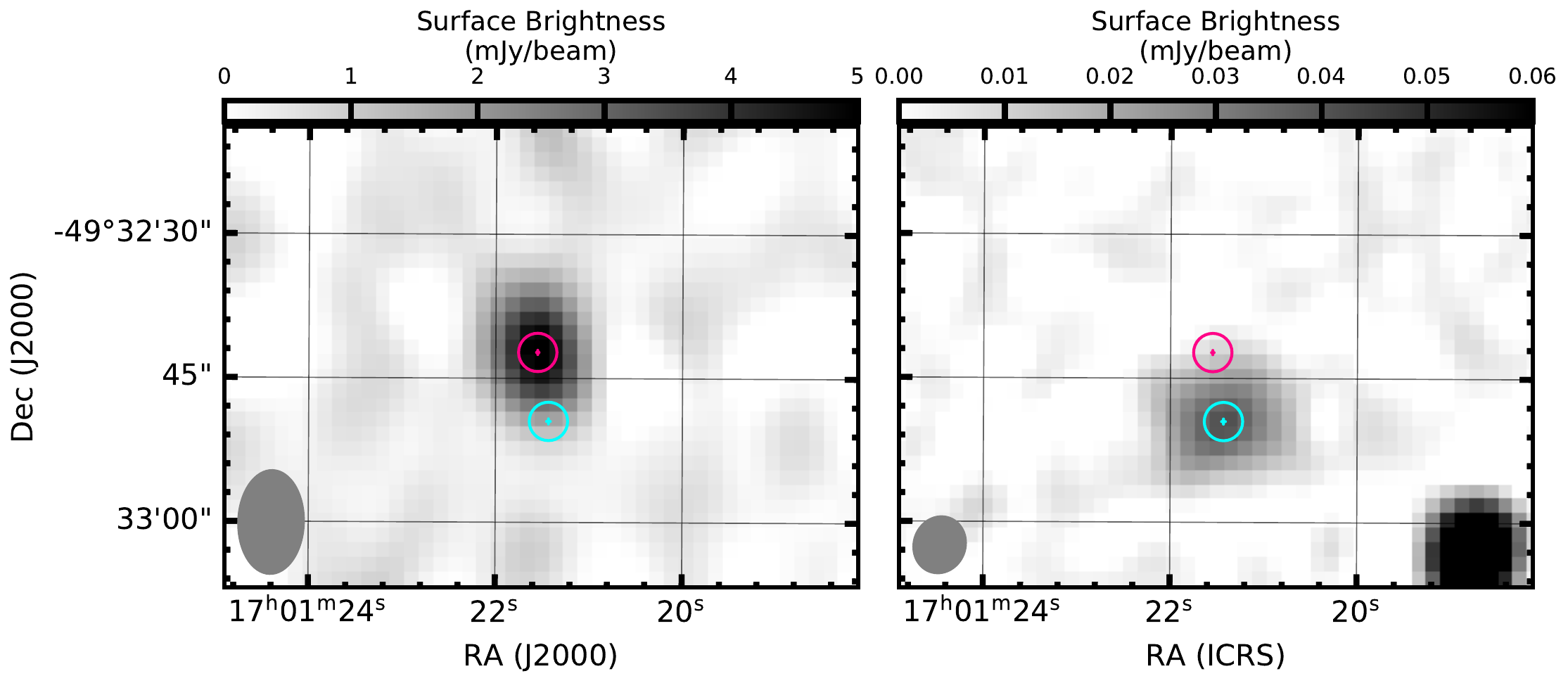}
\caption{MeerKAT images of the position of \mtpfrb. The left panel shows the 8\ s MeerKAT MFS image where \mtpfrb\ was detected and localised.
The position of \mtpfrb\ is marked with the magenta circle (2\arcsec radius) and cross hairs.
The right panel shows the 3.63\ hr combined image of the \gx\ field, with the location of the continuum source \prs\  marked with cyan circle (2\arcsec radius) and cross hairs. For both sets of cross hairs the size of the cross hairs reflects the sizes of the uncertainties on the positions. The synthesised beam is shown in the bottom left of each panel.}
\label{fig: MTP19 radio}
\end{figure*}

We applied the \trap\ to the 8\ s images. The resulting variability parameters are shown in Figure\ \ref{fig: MTP19 var params}. We can see that the known mode-changing pulsar, \gxpsr\ \citep{2007MNRAS.377.1383W,2019MNRAS.484.3691J}, is an outlier in all three parameters \citep[see][for the light curve of \gxpsr]{2022arXiv220309806D}. There is also another outlier in these plots that is more extreme in the \modp\ and \madp\ parameters compared to \gxpsr, and has a lower median flux density. This outlier was detected in a single 8\ s image, centred in time on 2021-04-05T04:14:42.2, which matches the time that MeerTRAP detected \mtpfrb. An image of the burst is shown in Figure\ \ref{fig: MTP19 radio} (left panel). As it was the only outlier in the variability parameters apart from known source \gxpsr, was detected in a single epoch, and was detected at the time of \mtpfrb, we identified this source as \mtpfrb.

We corrected the astrometry using the method described in Section\ \ref{sec: absolute astrometry}. We determined the transformation matrix, $A$, using the 10\ minute image:
\begin{align}
    A & = \begin{bmatrix}
    1.0 & -3.8\times10^{-3} & 1.0 \\
    8.5\times10^{-5} & 1.0 & 1.0 \\
    -1.2\times10^{-2} & 2.1\times10^{-3} & 1.0 \\
    \end{bmatrix}.
\end{align}
The separations between the MeerKAT sources and ATPMN sources before and after correcting the astrometry are shown in Table\ \ref{tab: astrometry sep}. We then applied the transformation matrix to the 8\ s image containing \mtpfrb. 
To confirm that the correction to the 10\ minute image was applicable to the 8\ s images we compared the positions of the bright sources in the field between the full integration time and 8\ s images. We found that the median offset of the positions was  0\farcs2, which is less than the 0\farcs4 uncertainty on the ATPMN positions. As such the uncertainty on the astrometry is the 0\farcs4 arcsecond absolute astrometric uncertainty from the ATPMN positions.
After applying the correction, the position of \mtpfrb\ was determined to be RA: 17$^{\mathrm{h}}$01${\mathrm{m}}$21\fhs5 (255.33970$^{\circ}$) and Dec: $-49^{\circ}$32\arcmin42\farcs5 ($-$49.54515$^{\circ}$). The uncertainty on the RA and Dec found by \pybdsf\  was 0\farcs1 and 0\farcs2 respectively; however, the uncertainty on the astrometric correction was 0\farcs4 in both RA and Dec. We added these uncertainties in quadrature to find that the uncertainty on the position of \mtpfrb\ is 0\farcs4 in RA and 0\farcs5 in Dec. 
The burst is 47\arcmin30\arcsec\ from \gx\  {and 47\arcmin\ 48\arcsec\ from the phase centre of the observation.}

\begin{table}
    \centering
    \caption{Separation between the ATPMN and MeerKAT reference sources before and after applying the transformation. The MeerKAT positions were extracted from the full, ten\ minute integration image. The separation is given in arcseconds.}
    \begin{tabular}{lrr}
    ATPMN source name & separation before (\asec) & separation after (\asec) \\
    \hline
    \hline
    J165418.2$-$481303 & 0.6 & 0.4 \\
    J165613.1$-$492318 & 0.6 & 0.2 \\
    J165614.9$-$472915 & 0.2 & 0.2 \\
    J165908.3$-$481548 & 0.1 & 0.1 \\
    J171154.9$-$491250 & 0.4 & 0.07 \\
    \hline
    \end{tabular}
    \label{tab: astrometry sep}
\end{table}

\subsection{Flux density and fluence}
\label{sec: flux density and fluence}

{\mtpfrb\ was detected 47\arcmin\ 48\arcsec\ from the phase centre of the observation. This means that we need to take the attenuation of the IB into account. We use the modified single-pulse radiometer equation \citep{1985Dewey} from \citet{2023arXiv230210107J} to do this:
\begin{equation}
    S_\text{peak} \left( \text{S/N}, \text{W}_\text{eq}, \vec{a} \right) = \text{S/N} \: \beta \: \eta_\text{b} \: \frac{ T_\text{sys} + T_\text{sky} }{ G \sqrt{b_\text{eff} N_\text{p} \text{W}_\text{eq}} } \: a_\text{CB}^{-1} \: a_\text{IB}^{-1},
    \label{eq:radiometer}
\end{equation}
where $S_\text{peak}$ is the peak flux density, 
$\text{S/N}=140.8$ is the signal-to-noise, 
$\beta=1$ is the digitisation factor, 
$\eta_{\mathrm{b}}=1$ is the beam-forming efficiency, 
$T_\text{sys}=19$\ K and $T_\text{sky}=7.8$\ K are the system and sky temperatures respectively, 
$G=0.335$ is the gain of the IB, 
$b_\text{eff}=670.4$\ MHz is the effective frequency bandwidth,
$N_\text{p}=2$ is the sum of the number of polarisations,
$\text{W}_\text{eq}=7.6\times10^{-3}$\ s is the observed equivalent width of the boxcar pulse, 
and {$a_\text{CB}^{-1}$=1 and $a_\text{IB}^{-1}=1/0.22$ are the CB and IB attenuation factors respectively that take into account the angular dependence of the beam response.} $a_\text{CB}^{-1}$=1 as we only consider the IB detection in detail here. Using these values results in $S_\text{peak}=15.9\ \mathrm{Jy}$ and a fluence of $F=120.8\ \mathrm{Jy\ ms}$. These values are shown in Table\ \ref{tab: mtp19 properties}.}






\subsection{Scattering}
\label{sec: scattering}

\begin{figure*}
  \centering
  \includegraphics[width=\columnwidth]{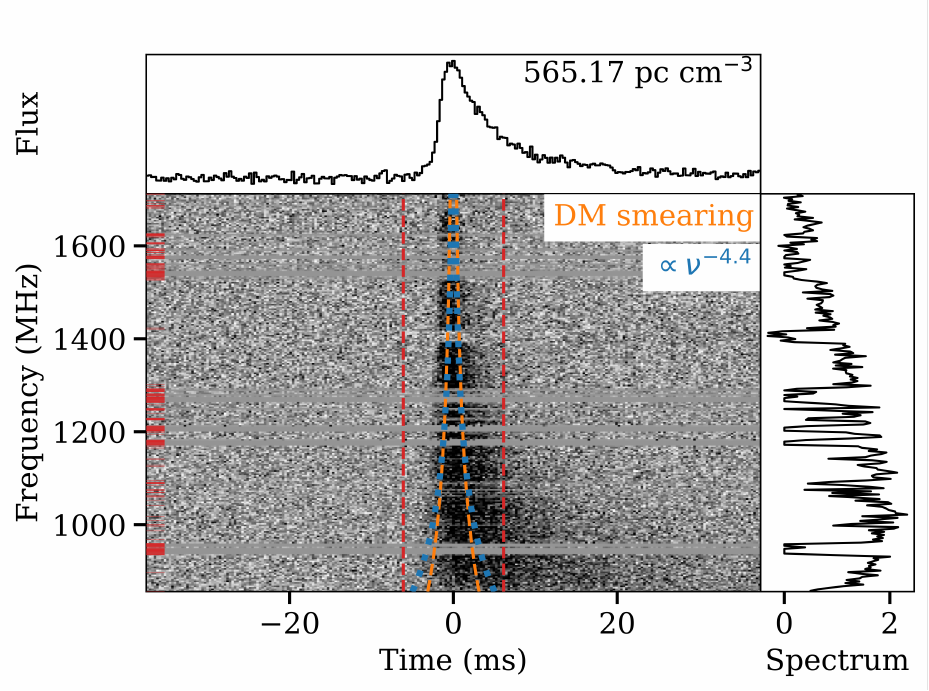}
  \includegraphics[width=\columnwidth]{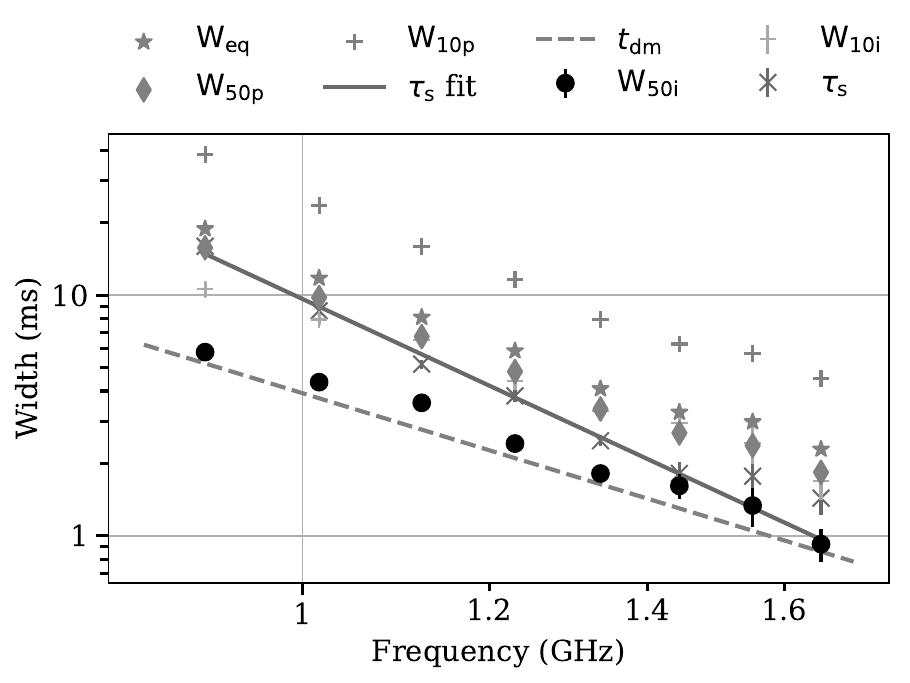}
  \includegraphics[width=\columnwidth]{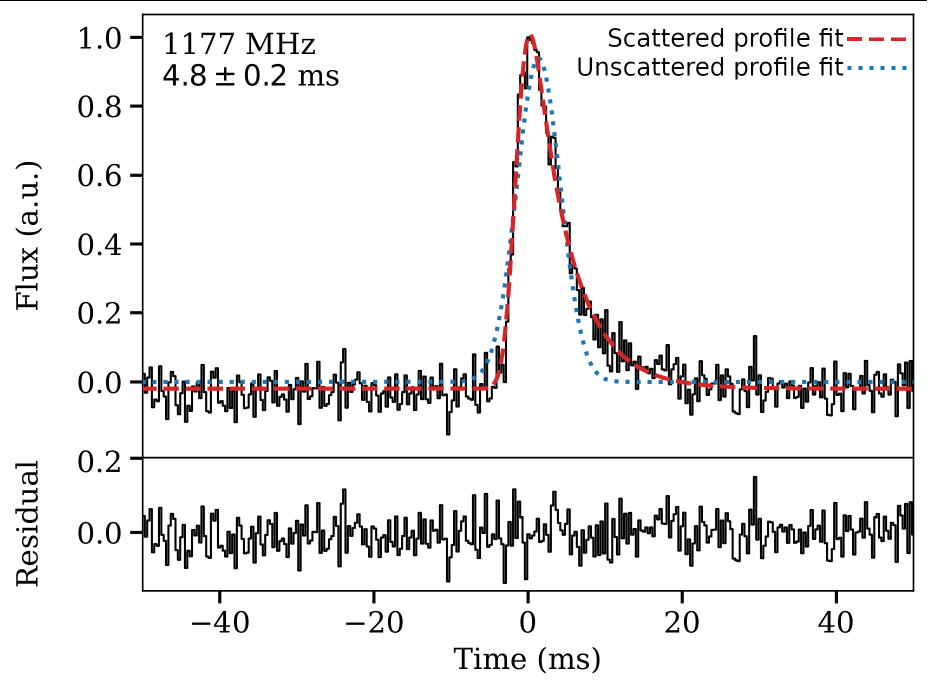}
  \includegraphics[width=\columnwidth]{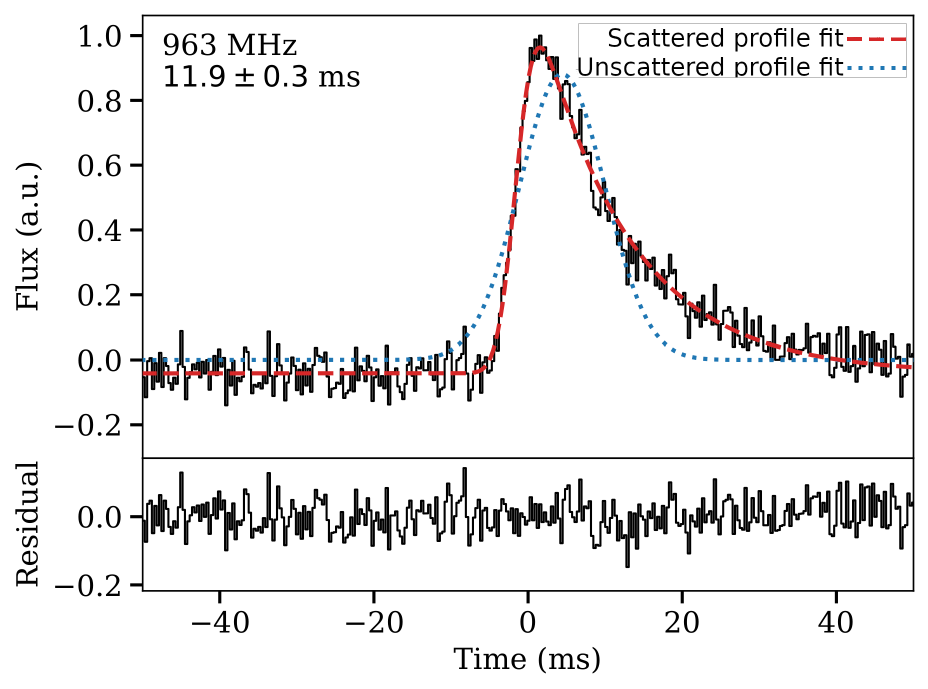}
  \caption{Scattering analysis of \mtpfrb. The MeerTRAP filterbank data were dedispersed at the scattering-corrected DM of $565.17~\text{pc}~\text{cm}^{-3}$. Top left: Dedispersed dynamic spectrum, pulse profile, and uncalibrated total intensity spectrum. The data are displayed at their native time resolution, but we summed every {four} frequency channels for clarity. The horizontal red lines indicate the masked channels, and the vertical dashed lines indicate the on-pulse region used to compute the spectrum. We used an {orange-dashed} line to mark the pulse broadening due to instrumental DM smearing. {We used a blue-dashed line to mark the pulse-broadening} $\propto \nu^{-4.4}$, where $\nu$ is the observing frequency, representative of scattering in Kolmogorov-turbulent ionised media. Top right: The post-scattering pulse widths $\text{W}_\text{50p}$ and $\text{W}_\text{10p}$, the Gaussian intrinsic widths $\text{W}_\text{50i}$ and $\text{W}_\text{10i}$, the boxcar equivalent widths $\text{W}_\text{eq}$, and the scattering times $\tau_s$ as a function of frequency. We show the expected DM smearing times (dashed line) and the best-fitting power law to the $\tau_s$ data (solid line). Bottom left: Pulse profile in a sub-band of 214~MHz centred at 1177~MHz with the best scattered profile fit (red dashed line) and the best unscattered Gaussian fit (blue dotted line) overlaid. The residuals are computed with respect to the scattering model and appear white. Bottom right: The same at 963~MHz.}
 \label{fig:scattering}
\end{figure*}

We analysed the highest-S/N total intensity pulse profile of \mtpfrb\  recorded by the MeerTRAP backend by fitting a scattering model to the frequency sub-banded data. The model consisted of a normalised Gaussian profile convolved with a single-sided exponential pulse broadening function that is characteristic for scattering in the thin-screen regime of turbulent ionised media (e.g.~\citealt{2001Cordes}). We performed the profile fits using a custom \textsc{python}-based software called \textsc{scatfit}\footnote{Version 0.2.14, \href{https://github.com/fjankowsk/scatfit/}{https://github.com/fjankowsk/scatfit/}} \citep{2022JankowskiScatfit}. It utilises the Levenberg–Marquardt minimization algorithm as implemented in \textsc{lmfit} \citep{2016Newville} for an initial fit, after which it explores the posterior using the \textsc{emcee} Markov chain Monte Carlo sampler \citep{2013ForemanMackey}. For more details about the fitting technique, see \citet{2023arXiv230210107J}. As the pulse broadening is significant, we first estimated the scattering-corrected DM from the data split into eight frequency sub-bands, which is $565.17 \pm 0.49~\text{pc}~\text{cm}^{-3}$. To determine the uncertainty we combined in quadrature the half width at which the S/N versus trial DM curve decreased by unity with the error in $\Delta \text{DM}$ derived from the scattering fit.
The best-fitting scattering index is $-4.6 \pm 0.1$, which we can see as the solid line in Figure\ \ref{fig:scattering}.


\subsection{Continuum radio emission}
\label{sec: persistent radio emission}

We searched the 3.63\ hr joint image (see Section\ \ref{sec: thunderkat imaging}) for continuum radio emission near the position of \mtpfrb. The joint image was astrometrically corrected using the method described in Section\ \ref{sec: thunderkat imaging} using the same sources shown in Table\ \ref{tab: astrometry sep}. {The transformation matrix for the joint image is:
\begin{align}
    A & = \begin{bmatrix}
    1.0 & 4.4\times10^{-5} & 1.0 \\
    -1.1\times10^{-4} & 1.0 & 1.0 \\
    -2.9\times10^{-3} & -1.2\times10^{-2} & 1.0 \\
    \end{bmatrix}.
\end{align}}
A faint, slightly resolved continuum source, \prs, is identified
7\farcs4 from the position of \mtpfrb, shown in Figure\ \ref{fig: MTP19 radio} (right panel).
{The uncorrected coordinates of the resolved source are RA: 17$^{\rm h}$01$^{\rm m}$21\fhs4s (255.33932$^{\circ}$) and 
Dec: $-$49$^{\circ}$32\arcmin50\farcs0 ($-$49.54708$^{\circ}$). The astrometrically corrected coordinates of \prs\ are RA: 17$^{\rm h}$01$^{\rm m}$21\fhs4s (255.33926$^{\circ}$) and 
Dec: $-$49$^{\circ}$32\arcmin50\farcs2 ($-$49.54727$^{\circ}$) with combined \pybdsf\ and astrometric uncertainty 
of 0\farcs6 in RA and 0\farcs5 in Dec. The offset between the uncorrected and corrected position is 0\farcs7.
{\prs\  is too faint to be detected in sub-band images, unlike \mtpfrb, where we can apply a primary beam correction. It is only detected  in the joint MFS image. The uncorrected \pybdsf\ total flux density of this source is $150\pm4\ \mathrm{\mu Jy}$.}}


\section{Optical observations}
\label{sec: optical obs}

\begin{figure}
\includegraphics[width=\columnwidth]{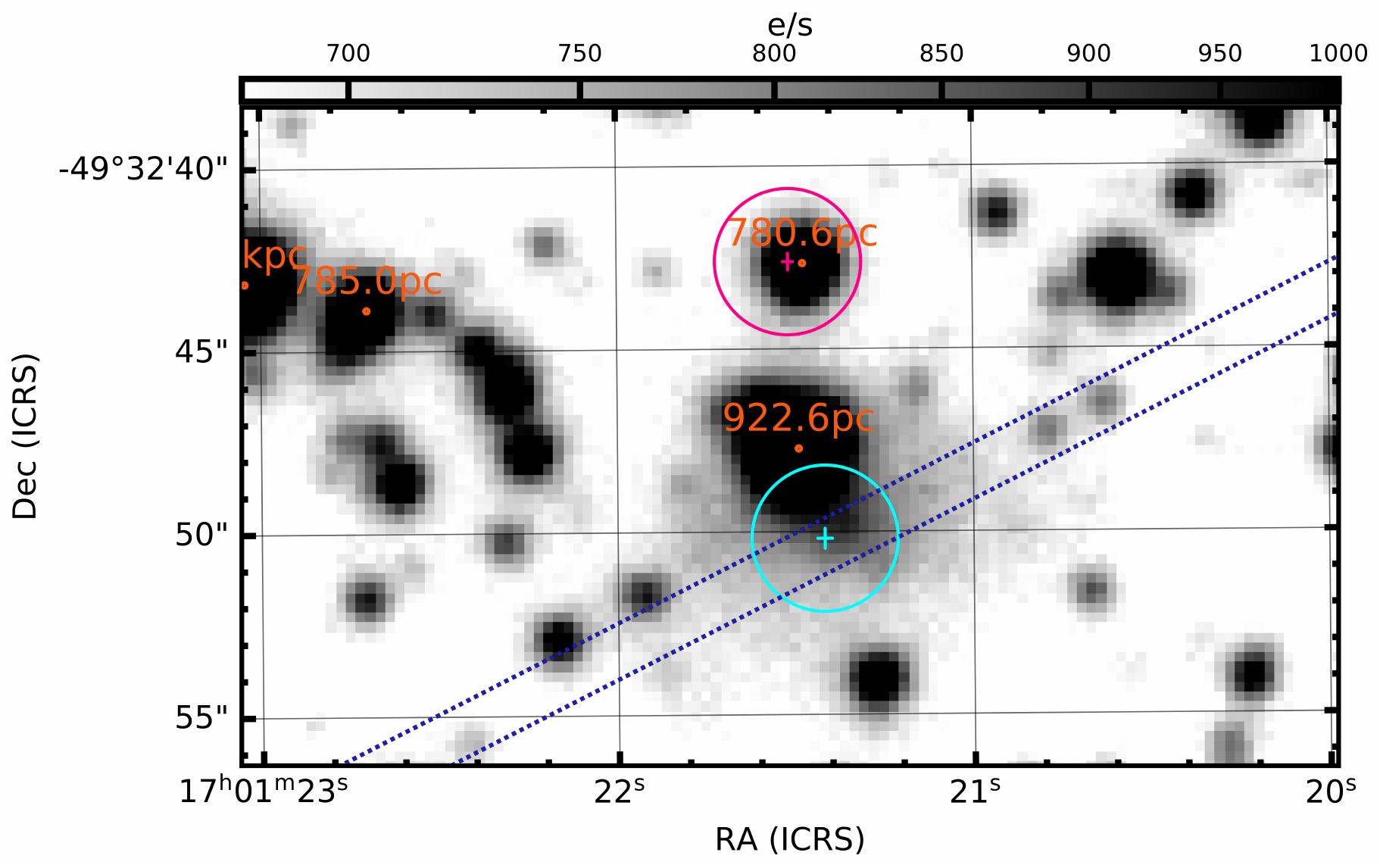}
\caption{DECaPS VR stacked optical image showing the astrometrically corrected positions of the \mtpfrb\ burst (magenta circle and cross hairs as in Figure\ \ref{fig: MTP19 radio}) and continuum radio source \prs\  (cyan  circle and cross hairs as in Figure\ \ref{fig: MTP19 radio}). Optical sources detected by the \textit{Gaia} DR3 mission with distance values, where available, are labelled in orange. {The blue dotted lines show the RSS slit position.}}
\label{fig: MTP19 DECaPS}
\end{figure}

A Dark Energy Camera Plane Survey \citep[DECaPS;][]{2018ApJS..234...39S} VR  stacked optical image is shown in Figure\ \ref{fig: MTP19 DECaPS} with the same magenta (\mtpfrb) and cyan (\prs) cross hairs as shown in Figure\ \ref{fig: MTP19 radio}. {The VR filter is a broadband filter from 497nm to 756nm. Stacked images are combined images from multiple exposures.}
Optical sources from {\textit{Gaia} Data Release 3 \citep[\textit{Gaia} DR3;][]{2016A&A...595A...1G,gaiadr3} that have \textit{Gaia} DR3 distances have also been labelled in Figure\ \ref{fig: MTP19 DECaPS}}. {\citet{2018ApJS..234...39S} found that the typical difference between DECaPs source positions and \textit{Gaia} source positions is 0\farcs1 and \textit{Gaia} is tied to the ICRS.}
Despite \mtpfrb\  being located in the Zone of Avoidance (ZoA), we can see a faint extended optical source coincident with the position of \prs. This suggests that the extended, continuum radio emission is associated with this extended optical source. {The ZoA is the part of the sky that is strongly affected by Galactic foreground extinction. In the ZoA the foreground extinction causes background galaxies to appear smaller as isophotal diameters are strongly affected \citep[see e.g.][]{2010MNRAS.401..924R}.}
We can also see in Figure\ \ref{fig: MTP19 DECaPS} that \mtpfrb\ is 0\farcs3 from a bright optical source, \hctwo, {that has a \textit{Gaia} DR3 distance of $780.6^{+80.2}_{-98.5}$\ pc \citep[][]{gaiadr3}.}

Optical spectroscopy of the extended optical source and \hctwo\  was undertaken with the Southern African Large Telescope \citep[SALT;][]{Buckley2006SPIE.6267E..0ZB} using the Robert Stobie Spectrograph \citep[RSS;][]{Burgh2003SPIE.4841.1463B}. 
Data reductions were done using {\tt PySALT} version 0.47, the PyRAF-based software package for SALT data reductions
\citep{Crawford2010SPIE.7737E..25C}\footnote{\href{https://astronomers.salt.ac.za/software/pysalt-documentation/}{https://astronomers.salt.ac.za/software/pysalt-documentation/}}, which includes gain and amplifier cross-talk corrections, bias subtraction, amplifier
mosaicing, and cosmetic corrections. Spectral reductions (object extraction, wavelength calibration and background subtraction) were all done using standard IRAF\footnote{\href{https://iraf.noao.edu/}{https://iraf.noao.edu/}} routines, including relative flux calibration.

Two repeat 1150 s exposures were taken of \hctwo\ 
on 2021 April 15 in clear conditions with seeing of 1\farcs8. The low resolution PG300 grating was used, covering the region 3800--8400\ \AA{}, with a slit width of 1\farcs25, resulting in a spectral resolution of 17\AA.  The combined spectrum appears to be a G or K type star, with an obvious Mg b line at 5172\AA~ and no presence of molecular bands, ruling out a later spectral type.

\begin{figure*}
\includegraphics[width=0.9\textwidth]{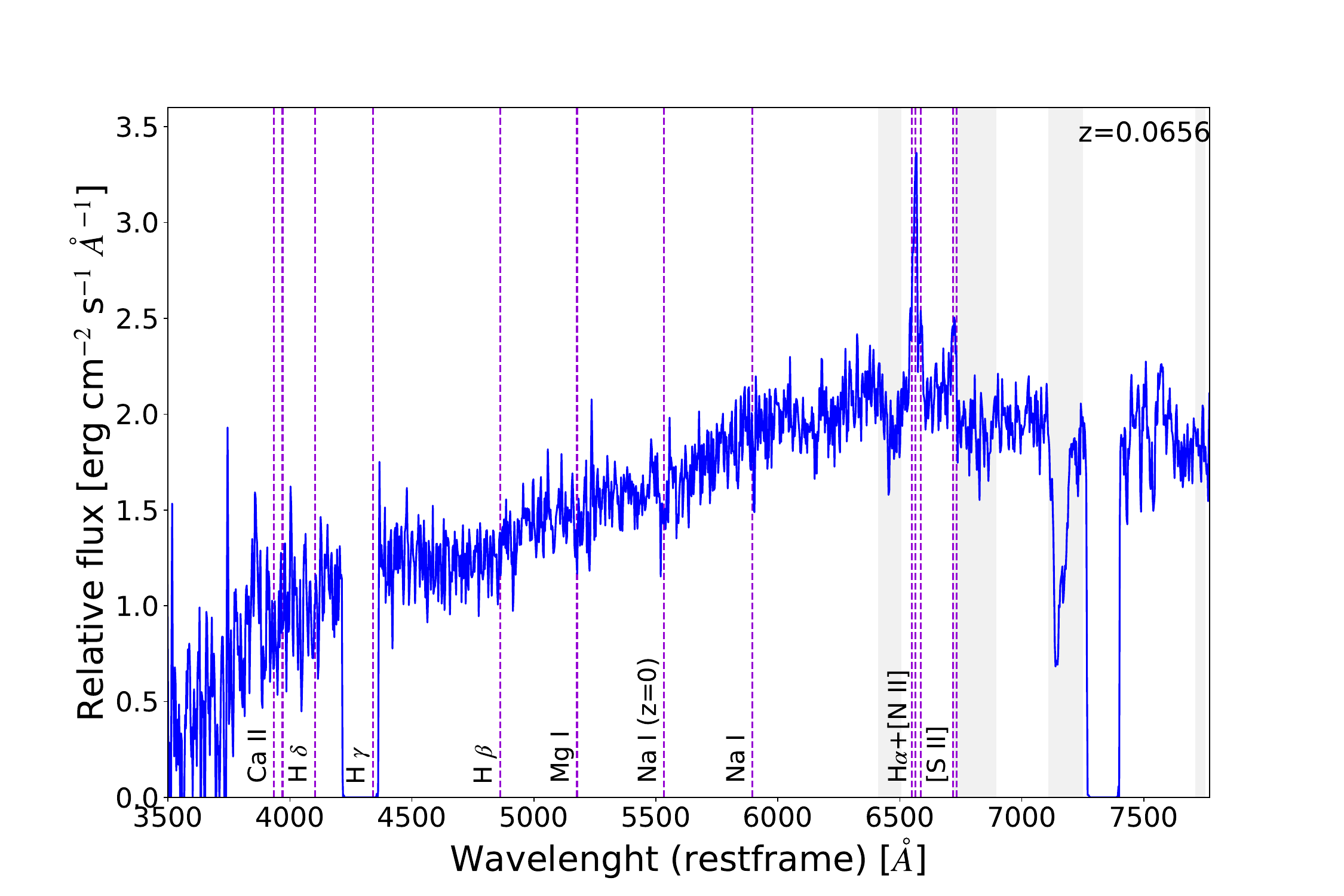}
\caption{SALT spectrum of the {{extended optical source}. Specific lines are indicated with dashed-purple lines, and telluric lines are highlighted in grey.}}
\label{fig: SALT-spectrum}
\end{figure*}

A SALT spectroscopic observation of the extended optical source coincident with \prs\  was obtained on 2021 September 02 in clear conditions with 1\farcs3 seeing using the same PG300 grating, but with a 2$^{\prime\prime}$ slit, giving a resolution of 24\AA.
The position of the slit, {chosen to reduce contamination from \hctwo, is shown in Figure\ \ref{fig: MTP19 DECaPS}. }
Two repeat 1600~s exposures were taken. This object is a galaxy, with prominent H$\alpha$ and [SII] emission lines at $z$ = 0.066 (see Figure~\ref{fig: SALT-spectrum}). The H$\alpha$ value suggests that it is consistent with a star-forming spiral galaxy. The disk of the galaxy is likely obscured by extinction \citep[$A_{V} = 2.41$ at the position of \mtpfrb;][]{2011ApJ...737..103S}.


\begin{figure*}
\includegraphics[width=\textwidth]{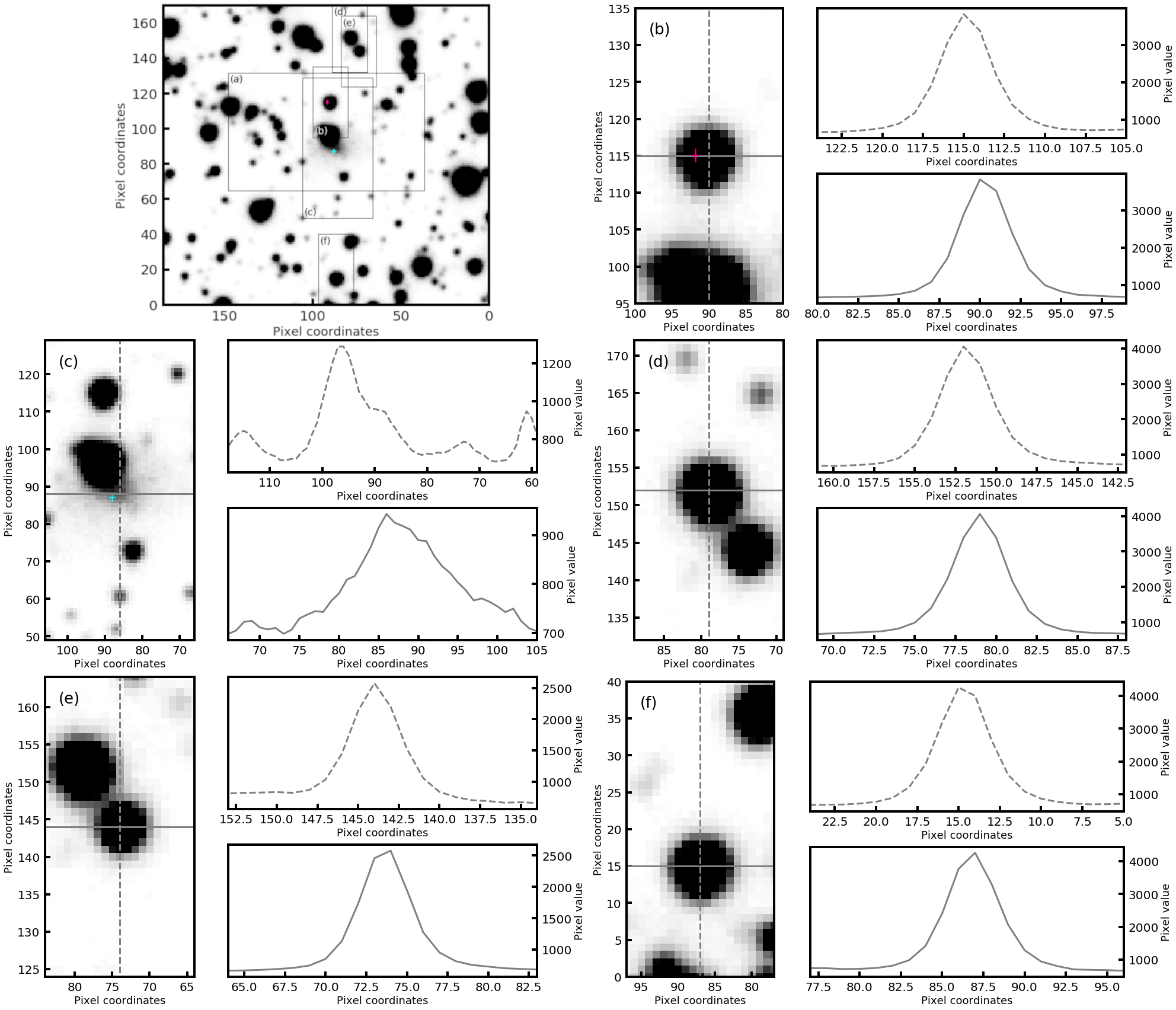}
\caption{DECaPS VR stacked optical image showing the PSFs of the optical sources near \mtpfrb. The top left panel shows the FoV around \mtpfrb\  and the positions of the nearby sources. The region marked (a) is the region shown in Figure\ \ref{fig: MTP19 DECaPS}. The other regions are shown in the other panels in this figure. {Region/panel} (b) includes the position of \mtpfrb\  (marked with magenta cross-hairs) and shows the cross-section of the PSF of \hctwo. {Region/panel} (c) shows the position of the galaxy, with the position of the continuum radio source marked with cyan cross-hairs. We can see that the optical PSF of the galaxy is wide and irregular, while the PSF of \hctwo\  and the nearby sources shown in panels (d) - (f) are approximately Gaussian, as expected for a point source.}
\label{fig: MTP19 DECaPS psfs}
\end{figure*}

We also examined the point spread function (PSF) of \hctwo, shown in Figure\ \ref{fig: MTP19 DECaPS psfs}(b), and compared it to the PSFs of other sources in the field, shown in panels (d) to (f) of Figure\ \ref{fig: MTP19 DECaPS psfs}. The PSFs of all of these sources show no evidence of excess emission in the wings and can be described as Gaussians.

\subsection{Properties of the extended optical source}
\label{sec: discussion properties of the host}

The optical galaxy is faint in the optical, visible as an extended blob in the stacked VR image shown in Figure\ \ref{fig: MTP19 DECaPS}. It is coincident with the continuum extended radio source \prs. It is also in the ZoA with high extinction and partially obscured by a bright foreground star. It is only because of the sensitivity of DECam that we could detect the galaxy at all in the optical. While the obscuring star and faintness make optical analysis challenging, we can investigate the source in other bands. An extended source is visible in the 2 Micron All Sky Survey \citep[2MASS;][]{2006AJ....131.1163S}, particularly in the {K band. The J and H bands} are likely dominated by contamination from the nearby star. It is identified as a galaxy, \irgal, and has an extinction corrected ($A_{K}=0.265$) apparent magnitude of $\sim14.4$. Using the SALT distance, this corresponds to an absolute K-band magnitude of $\sim-22.8$, which is also consistent with a spiral galaxy. The galaxy is visible as an extended source in the WISE W1 band, with WISE colours $\mathrm{W1}-\mathrm{W2}\sim-0.169$ and $\mathrm{W2}-\mathrm{W3}\sim3.67$. This places it between where normal disk galaxies and star-forming galaxies are located in the WISE color-color diagram \citep[Figure\ 12 in][]{2010AJ....140.1868W}. This agrees with the SALT H$\alpha$ detection. 

Using the SALT spectrum, we determined the $\rm H\alpha$ line flux by simultaneously fitting $\rm H\alpha$ and the surrounding $\rm N\ {\sc II}$ doublet with three Gaussian profiles to the continuum-subtracted spectrum, excluding regions of sky absorption.
We obtained $F_{\rm H\alpha}=(1.9\pm0.1)\times10^{-16}$\ erg\ s$^{-1}$\ cm$^{-2}$, which corresponds to a star-formation rate limit {(SFR)} $\gtrsim 0.02\ \rm M_\odot$\ yr$^{-1}$ at $z = 0.066$ (SFR $\gtrsim 0.1\ \rm M_\odot$\ yr$^{-1}$ correcting for the Galactic extinction), considering a \citet{1955ApJ...121..161S} Initial Mass Function (IMF).
{These values should also be corrected for the dust extinction of the FRB host galaxy. We estimated this by measuring the flux limit of the $\rm H\beta$ flux and comparing it to the $\rm H\alpha$ over $\rm H\beta$ Balmer ratio. We found significant dust extinction with A$_V>1.8$. We corrected the $\rm H\alpha$ flux for dust extinction in the host galaxy using a Milky Way extinction curve by \cite{Pei1992} and obtained $F_{\rm H\alpha}>(3.4\pm0.1)\times10^{-15}$\ erg\ s$^{-1}$\ cm$^{-2}$, corresponding to an SFR $\gtrsim 0.3\ \rm M_\odot$\ yr$^{-1}$.}
Using the absolute K band magnitude $\sim-22.8$ and the mass-luminosity relation of galaxies, we estimated a stellar mass of $\rm{log(M_*/M_{\odot})} = 11.25$.
{Using this value and the dust corrected SFR determined above, we obtained a specific star formation rate limit (sSFR, defined as SFR/$\rm{M_*}$) of
$\log(\rm sSFR) > -11.7$ yr$^{-1}$.}
Finally, using the flux of the $\rm N\ {\sc II}\ \lambda6584$, we estimated the metallicity 12 + log(O/H) = $9.0^{+0.2}_{-0.1}$, using the strong line ratio with $\rm H\alpha$ and following \citet{2008A&A...488..463M}.

We used a two-hour MeerKAT observation of the \gx\  field from 2021 April 07 to search for HI emission from \prs. Using the redshift of the extended optical source, $z = 0.066$, we find that the expected HI frequency is 1.3324~GHz; however, we do not detect any HI emission. This is likely because the galaxy lies nearly a degree from the phase centre of the MeerKAT observations, which is outside the primary beam and is therefore challenging to detect.
The root mean square (RMS) noise of the HI cube is 0.23\ mJy\ beam$^{-1}$\ channel$^{-1}$ prior to primary beam correction and  1.15\ mJy\ beam$^{-1}$\ channel$^{-1}$ after primary beam correction. The beam size in the HI cube is $16\farcs45 \times7\farcs00 \sim 20.8\times8.9$ kpc.
We derived the 3$\sigma$ upper limit on the HI flux of the likely host galaxy using:
\begin{equation}
S_{\rm HI\_ul} = 3 \times rms \times W \times \sqrt{\frac{W_0}{W}}
\ ,
\end{equation}
where $S_{\rm HI\_ul}$ is in the units of ${\rm Jy\ km\ s^{-1}}$, rms is in units of Jy beam$^{-1}$ channel$^{-1}$ for the HI cube after primary beam correction, $W$ is the velocity width of the HI emission for the likely host galaxy in units of ${\rm km\ s^{-1}}$, $W_0$ is 44.5 ${\rm km\ s^{-1}}$ (the channel width of the HI data cube). By assuming $W$ is 400 ${\rm km\ s^{-1}}$ and the HI disk is smaller than the beam size ($20.8\times8.9$ kpc), $S_{\rm HI\_ul}$ is 0.46 Jy ${\rm km\ s^{-1}}$.
The HI mass upper limit for the likely host galaxy is $8.4\times10^9\ \rm M_\odot$, given by \citet{Meyer2017PASA...34...52M}:
\begin{equation}
M_{\rm HI\_ul} = \frac{2.35 \times 10^5}{(1+z)^2} \times D_L^2 \times S_{\rm HI\_ul}
,
\end{equation}
where $M_{\rm HI\_ul}$ is in M$_\odot$, $S_{\rm HI}$ is the HI flux upper limit in ${\rm Jy\ km\ s^{-1}}$, $z$ is the redshift (0.066) and $D_L$ is the luminosity distance of the galaxy in Mpc (296.8 Mpc is used for $z = 0.066$ and assuming ${\rm H_0 = 70 \: km\ s^{-1}\ Mpc^{-1}, \Omega_M = 0.3, \Omega_{vac} = 0.7}$).






\section{Discussion}
\label{sec: discussion}

\subsection{Is \mtpfrb\ Galactic or extra-galactic?}
\label{sec: is it galactic}

We have detected and localised \mtpfrb~using the MeerTRAP pipeline and ThunderKAT images. However, there is a discrepancy between the {Milky Way DM contributions from the \textsc{ymw16} model (348.7\ \pcc) and the \textsc{ne2001} model (516.1\ \pcc)}, that may place \mtpfrb\  ($\sim566$\ \pcc) either just outside the Galaxy, or inside the halo. Here we will present the properties of \mtpfrb\ to determine if the source is Galactic or if it originates from the identified candidate galaxy: \irgal.

\subsubsection{Nearby star \hctwo}
\label{sec: discussion nearby star}

{The \textit{Gaia} DR3 distance of \hctwo\  is {$780^{+80}_{-100}$}\ pc. Using the \textsc{ymw16} electron density model,
a source in that direction within the Milky Way Galaxy at that distance would have a DM of $\sim24$\ pc\ cm$^{\mathrm{-3}}$
compared to the DM of \mtpfrb: $\sim566$\ \pcc. {The PSF of \hctwo\ is a Gaussian that matches the PSF of point sources in the field. This indicates that the source is a point source as expected for a star and that there is no evidence of a galaxy directly behind and obscured by \hctwo.}
Due to the spectral confirmation of \hctwo\  as a star, the distance to it, and the lack of evidence of an extended source {directly behind the star} in the PSF, we conclude that the position match between \mtpfrb\ and \hctwo\  is a chance coincidence. }

{\hctwo\ is 0\farcs3 from the position of \mtpfrb. The star's coordinates, after accounting for proper motion, are RA: 17$^{\mathrm{h}}$01${\mathrm{m}}$21\fhs477$\pm1\,\mathrm{mas}$ and Dec: $-49^{\circ}$32\arcmin42\farcs688$\pm1\,\mathrm{mas}$. The median offset between \textit{Gaia} counterparts of the third realisation of the International Celestial Reference Frame \citep[ICRF3][]{2020A&A...644A.159C} sources is 0.5\,mas \citep{2022A&A...667A.148G}. While the uncertainty on the star's position is small, the uncertainty on \mtpfrb's position is 0\farcs5 in RA and 0\farcs4 in Dec (taking the astrometric uncertainty into account). 
\hctwo\ is a G or K type star, so we can assume that it is a similar star to our Sun. Using the equations in \citet{tiburzi2021}, the radius at which the solar wind produces a DM of 60\ \pcc, accounting for the difference between the \mtpfrb\ DM and the \textsc{ne2001} MW DM, is 0.1\ AU. At a distance of {$780^{+80}_{-100}$}\ pc this is a disk of radius $0.13^{+0.02}_{-0.01}\,\mathrm{mas}$. This is significantly smaller than the uncertainty on the star's position. Even if we assume that \hctwo\ could produce significant DM contributions at 1\ AU the radius of the disk at {$780^{+80}_{-100}$}\ pc would be $1.3^{+0.2}_{-0.1}\,\mathrm{mas}$.
Given that the radius of the possible stellar wind is at least two orders of magnitude smaller than the uncertainty on the FRB position it is unlikely that \mtpfrb\ passed close enough to the star for any possible stellar winds to have a significant impact on the DM of the FRB.}


\subsubsection{Dispersion measure}
\label{sec: discussion DM}

We can use the DM to explore whether the FRB progenitor is likely in the galaxy or in the Milky Way Halo. We use {a halo} DM of {$144\pm60$\ \pcc}\  from \citet{2020Yamasaki} including a 40\ per\ cent uncertainty. This is because \mtpfrb\ is at a Galactic latitude of $b=-4.597$~deg or $\sin  \left| b \right|  = 0.08$. Constraining the DM of the halo, particularly near the Galactic plane, is challenging. Recently, \citet{2023arXiv230103502C} presented a comparison of halo DM models and reviewed them using FRB DMs. While they only presented models with $\sin  \left| b \right|  > 0.5$, they found that the \citet{2020Yamasaki} halo model is consistent with their FRB observations.

\begin{figure}
  \centering
  \includegraphics[width=\columnwidth]{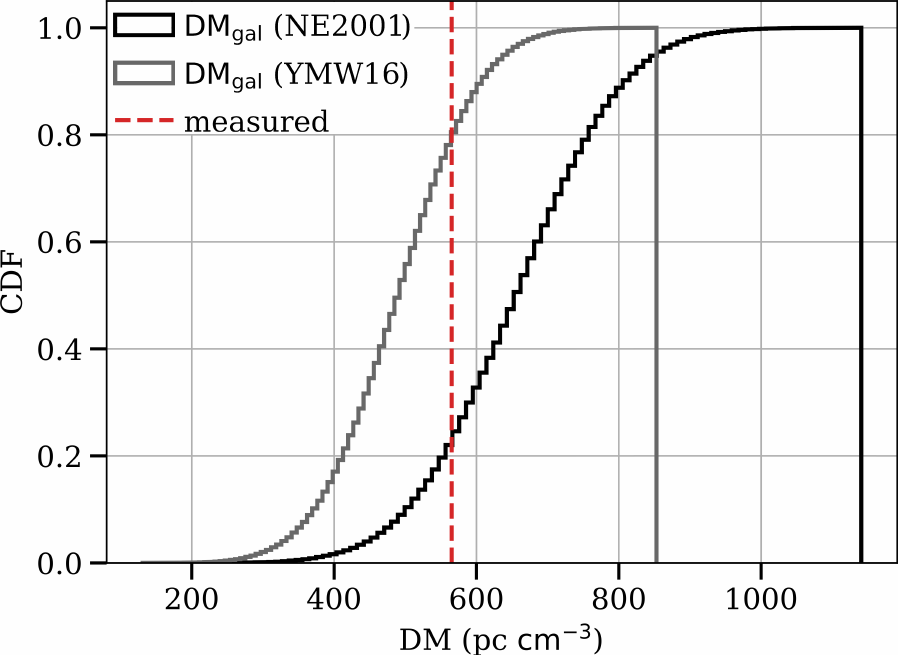}
  \caption{{Cumulative density function of the total Galactic DM contribution $\text{DM}_\text{gal}$ = $\text{DM}_\text{ism}$ + $\text{DM}_\text{halo}$ for the \textsc{ne2001} and \textsc{ymw16} model and using the \citet{2020Yamasaki} halo model. We assumed a 20~per cent uncertainty on the maximum integrated ISM contributions and a 40~per cent uncertainty on the MW halo component. The red vertical line marks \mtpfrb's measured DM. The probabilities of \mtpfrb\  being extra-galactic are $\sim$24 and $\sim$80~per cent, respectively.}}
 \label{fig:distanceprobability}
\end{figure}

The ISM contributions computed using the \textsc{ne2001} and \textsc{ymw16} Galactic free electron models differ significantly. If we assume a 20~per cent uncertainty on the maximum integrated ISM DM contribution and a 40~per cent uncertainty on the MW halo DM contribution (the halo is more poorly understood) for the \mtpfrb\  sightline, the probabilities of \mtpfrb\  being extra-galactic are $\sim$24 and $\sim$80~per cent for the \textsc{ne2001} and the \textsc{ymw16} model, respectively. We visualise this in Figure~\ref{fig:distanceprobability}.

We can use the DM inventory to consider whether the different DM contributions are feasible. 
The \textsc{ymw16} DM excess is {$\mathrm{DM_{FRB}} - \mathrm{DM_{MW,\textsc{ymw16}}} = 217$} \pcc\ and the \textsc{ne2001} DM excess is {$\mathrm{DM_{FRB}} - \mathrm{DM_{MW,\textsc{ne2001}}} = 50$}\ \pcc. We assume that the uncertainty on these DMs is 20\ per\ cent. The expected cosmic DM for $z=0.06$ is $\mathrm{DM_{cosmic}} = 51^{+82}_{-25}$\ \pcc using the Macquart relation \citep{2022MNRAS.509.4775J}.
If we use the \textsc{ymw16} DM and assume that \mtpfrb\ is associated with the galaxy at $z=0.066$ then $\mathrm{DM_{host}} = \mathrm{DM_{FRB}} - \mathrm{DM_{halo}} - \mathrm{DM_{ISM}} - \mathrm{DM_{cosmic}} = 565.17 - 144\pm58 - 349\pm70 - 51^{+82}_{-25}  = 21^{+122}_{-94}$\ \pcc. Assuming the same values but using \textsc{ne2001} for the $\mathrm{DM_{ISM}}$ gives $\mathrm{DM_{host}} = -145^{+144}_{-121}$\ \pcc.
The uncertainties on these host DM values are large as the DMs along the line of sight are not well known. 
{The $1\sigma$ upper limit on the \textsc{ne2001} $\mathrm{DM_{ISM}}$ is -1\ \pcc; however, the range of possible $\mathrm{DM_{host}}$ values using \textsc{ymw16} for $\mathrm{DM_{ISM}}$ is plausible.}

However, we would require an impossible, negative $\mathrm{DM_{host}}$ if we use the \textsc{ne2001} model for $\mathrm{DM_{ISM}}$.

We need to account for DM excesses of $217\pm58$\ \pcc\  and $50\pm103$\ \pcc\  for \textsc{ymw16}  and \textsc{ne2001} respectively if we assume that \mtpfrb\ is inside the Milky Way. An unidentified HII region could account for these DM excesses; therefore, it is not possible to conclusively rule out that \mtpfrb\ originated from inside the Galaxy based on the DM alone.


\subsubsection{Scattering}
\label{sec: discussion scattering}

\begin{figure}
  \centering
  \includegraphics[width=\columnwidth]{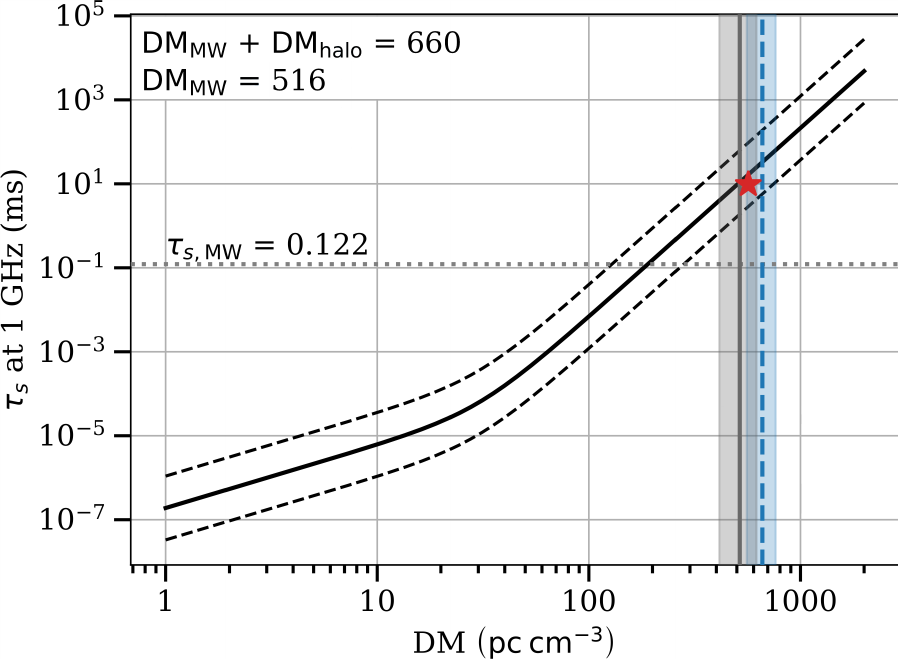}
  \caption{Location of \mtpfrb\  (red star) in a scattering time, $\tau_s$, versus DM diagram. {$\tau_{\mathrm{s, MW}}$ is the expected Milky Way scattering contribution.} We highlight the $\tau_s$ -- DM relation of Galactic pulsars and its 1-$\sigma$ uncertainty range with black lines. We also show the DM contributions and their uncertainty ranges from the ISM and the MW halo for that sightline, assuming the \textsc{ne2001} free-electron model and the \citet{2020Yamasaki} halo model. }
 \label{fig:distance}
\end{figure}

Figure~\ref{fig:distance} shows the scattering time, $\tau_s$, measurement of \mtpfrb\ interpolated to 1~GHz in a $\tau_s$ -- DM diagram. We highlight the expected DM contribution from the Milky Way (MW), $\text{DM}_\text{MW} = 516 \pm 103~\text{pc}~\text{cm}^{-3}$, with its 20~per cent uncertainty range for that sightline as computed using the \textsc{ne2001} Galactic free-electron model. Additionally, we show the combined range of $\text{DM}_\text{MW}$ and the MW halo DM contribution of $144~\text{pc}~\text{cm}^{-3}$ expected from the \citet{2020Yamasaki} halo model. The curved black solid line shows the best-fitting $\tau_s$ -- DM relation of Galactic pulsars \citep{2022Cordes} with its 1-$\sigma$ range highlighted with black dashed lines.
{We can see in Figure\ 9 that \mtpfrb's scattering time is larger than the scattering time expected from the Milky Way, {which is also the lower DM limit for an extra-galactic source}; however, it is consistent with the scattering-DM relation of Galactic pulsars. This implies that
\mtpfrb\ 's scattering time measurement is consistent with either an extra-galactic source or a source within the MW halo.}

\mtpfrb\  is located at a low Galactic latitude of $-4.597$~deg, which means that its radio signal passed through the thick disk of the Milky Way. The best-fitting scattering index is $-4.6 \pm 0.1$ (see Section\ \ref{sec: scattering}), which agrees well with the power law index expected from Kolmogorov turbulence in ionised media, $-22/5 = -4.4$. It is also consistent with the spread in scattering indices $3 < \gamma_s < 4.5$ observed in Galactic radio pulsars \citep{2021Oswald, 2022Cordes}. The scattering time interpolated to 1~GHz is $9.7 \pm 0.2~\text{ms}$. The pulse broadening expected from the Milky Way ISM in that direction is $0.122$~ms according to the \textsc{ne2001} Galactic free-electron model \citep{2002Cordes}. The isotropic thin-screen scattering model describes the data well with profile residuals that appear normally distributed for both the band-integrated and sub-banded data. The profile data offer little room for deviations from the isotropic scattering regime.


\subsubsection{Luminosity}
\label{sec: discussion luminosity}

\begin{figure}
\includegraphics[width=\columnwidth]{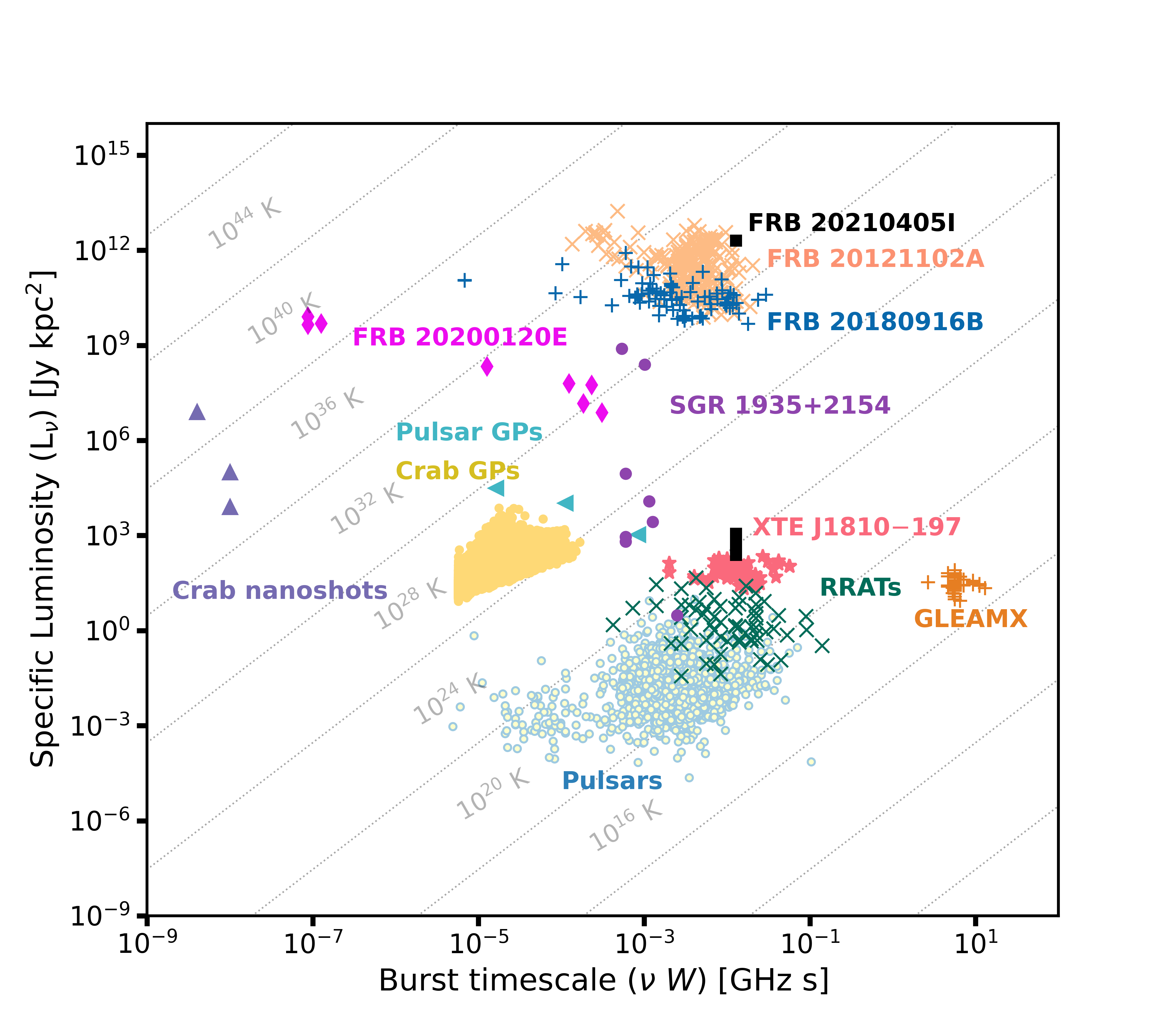}
\caption{Specific luminosity versus burst duration plot for various coherently emitting radio transients \citep[][]{2003ApJ...596.1142C,2022NatAs...6..393N}. Grey lines indicate lines of constant brightness temperature. \mtpfrb\ is marked in black {twice in the plot, once assuming it is extra-galactic and once assuming it is Galactic}. 
}
\label{fig: MTP19 energetics}
\end{figure}

We investigated the expected specific radio luminosity of~\mtpfrb~to verify whether the expected energetics are consistent with the general population of FRBs. To do so, we computed the specific luminosity by multiplying the peak flux density of the burst with the square of the distance to the source. We consider two cases: 1) \mtpfrb~lies within our Galaxy and 2) \mtpfrb~lies in the galaxy at a redshift of 0.066. In order to estimate the distance of \mtpfrb~in the Galaxy, we used a DM range of 416 to 616~pc~cm$^{-3}$ and use the \textsc{ne2001} electron density model to map the DM to a distance in the Galaxy. Then we computed the specific luminosity for \mtpfrb~for this range of distances. The values for both cases were plotted on the specific luminosity versus burst width parameter space that shows different classes of coherently emitting transients (see Figure~\ref{fig: MTP19 energetics}). One can clearly see from the figure that if \mtpfrb~is associated with the optical galaxy at a redshit of 0.066 then the luminosity is consistent with the population of FRBs. 
{On the other hand, if \mtpfrb~lies in our Galaxy, the burst has a higher specific luminosity than bursts from the magnetar XTE\ J$1810-197$ and stands out from the populations of RRATs, magnetars, pulsars {based on luminosity} {and stands out from} pulsar giant pulses {based on timescale}.}

\subsubsection{Association with \irgal}
\label{sec: discussion prs chance coincidence}

{In order to robustly claim association of \mtpfrb~with the \irgal~, we computed the chance coincidence probability of spatial coincidence. For this we considered two cases separately: 1) \mtpfrb~is Galactic and lies on top of \irgal~in the sky plane and 2) \mtpfrb~is extra-galactic but originates in a different unseen host galaxy. For case 1, the probability of chance coincidence,
\begin{equation}
    P(cc| l,b) = P(\mathrm{2MASS\ J1701249-4932475}, l, b),
\end{equation}
where, $P(\mathrm{2MASS\ J1701249-4932475}, l, b)$ is the probability for a galaxy to be at a Galactic longitude $l$ and latitude $b$. To obtain $P(\mathrm{2MASS\ J1701249-4932475}, l, b)$ out to the redshift of \irgal, we used the same methodology as used by~\cite{2008MNRAS.391..935C} and found a probability of 0.0025. 
Hence, for case 1, $P_{cc}\simeq$~2.5$\times$10$^{-3}$.}

{For case 2, we used the chance association probability methodology from~\cite{2017ApJ...849..162E} and \cite{2021ApJ...911...95A}. For this case, we used the 99$\%$ confidence upper limit of an unseen host redshift of 0.35.
Using this, we obtained $P_{cc}\simeq$0.009. So for both cases, the value of $P_{cc}$ is very low. Therefore, we conclude that \mtpfrb~is associated with \irgal.}

{The angular offset between \mtpfrb\  and the centre of the galaxy of 7\farcs4 at a luminosity distance of 296.8~Mpc corresponds to a projected physical offset of $10.65 $~kpc in the host galaxy reference frame. This places the FRB in the spiral arms or halo of the galaxy. This is at the high end of the offset distribution of well-localised FRBs and clearly above the median of 3.3~kpc \citep{2020ApJ...903..152H} but is consistent with two of the bursts, FRB~20190611B ($z$ = 0.3778) and FRB~20191001A ($z$ = 0.2340).}

\subsubsection{Summary}
\label{sec: discussion summary of eg/G arguments}

{Neither the DM nor scattering properties of \mtpfrb\ strongly support either a Galactic origin or an association with \irgal\ at $z=0.066$. However, the specific luminosity favours the latter. We also show that the probability that the spatial coincidence between \mtpfrb\ and \irgal\ is chance coincidence is low.
Therefore, we find it much more likely that \mtpfrb\ is an FRB associated with \irgal\ as the DM, scattering, specific luminosity and localisation are all consistent with this conclusion. It is still possible for the FRB to originate within the Milky Way; however, this would require an unusually bright magnetar/pulsar burst and an improbable spatial coincidence.}

{The uncertainty in the Galactic or extra-galactic origin of this FRB highlights the uncertainty in our understanding of the Milky Way DM and particularly the Milky Way halo DM. This was also highlighted by \citet{2023arXiv230101000R} for FRB~20220319D. It is important to continue to test and expand our measurements of these as demonstrated by \citet{2023arXiv230103502C}. We are currently limited in our understanding and models in the Galactic Plane for both the Galactic and halo DM contributions. More FRB detections at lower Galactic latitudes will help to test and improve the models.}


\subsection{Comparison to previously identified FRB host galaxies}
\label{sec: discussion frb host galaxies}

These results indicate that \irgal\ is an evolved massive galaxy with high-metallicity as is typical at such high stellar masses, but with very low star-formation. { This possibly points towards even higher dust extinction in the host galaxy or to an early type spiral galaxy. The stellar mass and SFR are respectively at the higher and lower end} of the FRB host population \citep{2022AJ....163...69B}, extending it to massive, early type galaxies. A comparison of these characteristics with the hosts of other classes of transients, shows that these properties are completely different from those of the hosts of explosive transients such as long gamma-ray bursts \citep{2017A&A...599A.120V,2019A&A...623A..26P} and superluminous supernov\ae\  \citep{2016A&A...593A.115J,2014ApJ...787..138L}. However, they can correspond to a subgroup of the host galaxy population of other kinds of core-collapse SN (such as  Type II SN \citep{2014ApJ...787..138L,2014ApJ...789...23K}), if the real SFR is much higher than the reported lower limit. Some hosts of merger-driven transients such as short gamma-ray bursts \citep{2014ARA&A..52...43B} have similar characteristics (see also \citealt{2022AJ....163...69B}).

\subsection{Searching for \mtpfrb\ in past images}
\label{sec: Searching for mtpfrb in past images}

While \mtpfrb\ was detected by MeerTRAP and ThunderKAT on 2021 April 05, ThunderKAT has been observing \gx\ weekly since September 2018 and will continue to do so until at least August 2023. MeerTRAP started observing commensally with ThunderKAT in September 2021. This means that there are $\sim2$ years of weekly images prior to MeerTRAP observing with ThunderKAT where \mtpfrb\ may be detected if it is repeating and bright enough to be detected in 8\ s images. Producing thousands of 8\ s images is time-consuming, and not all of the available weekly epochs have been imaged in 8\ s slices. The list of epochs and the number of 8\ s images available for each epoch is shown in Table\ \ref{tab: ThunderKAT 8 second images} in Appendix\ \ref{appendix}. We have searched 100 epochs with a total of $8049\times8$\ s images observed prior to September 2021 for a repeat burst from \mtpfrb, using a match radius of 3\arcsec. We use a radius of 3\arcsec\ as the astrometric offset of each image can be different prior to correction.
We present some possible detections with low S/N in Table\ \ref{tab: possible repeats}. 

\begin{table}
    \centering
    \caption{Low S/N detections of sources in 8 second images that are within 3\arcsec\ of \mtpfrb, the first row shows the detection of \mtpfrb\ on 2021 April 05. The positions quoted here have been astrometrically corrected and the separation is the angular separation between the source and the \mtpfrb\ position. $S_{\nu}$ is the MFS (1283 MHz) peak flux density that has not had a primary beam correction applied.}
    \begin{tabular}{lA{2.5cm}rrr}
    Date & Coordinates (J2000) & Sep (\arcsec) & S/N & $S_{\nu}$ (mJy) \\
    \hline
    \hline
    2021.04.05 & 17$^{\mathrm{h}}01^{\mathrm{m}}21\fhs5\pm0\farcs4$ $-49^{\circ}32\arcmin42\farcs5\pm0\farcs5$ & 0   & 25 & $5.2 \pm 0.2$\\
    2019.08.04 & $17^{\mathrm{h}}01^{\mathrm{m}}21\fhs5\pm0\farcs8$ $-49^{\circ}32\arcmin41\farcs5\pm2\farcs0$ & 1.0 & 4  & $0.9 \pm 0.2$\\
    2020.04.18 & $17^{\mathrm{h}}01^{\mathrm{m}}21\fhs7\pm0\farcs9$ $-49^{\circ}32\arcmin43\farcs2\pm2\farcs0$ & 1.9 & 4  & $0.7 \pm 0.2$\\
    \hline
    \end{tabular}
    \label{tab: possible repeats}
\end{table}

At the time of writing there were $\sim100$ weekly ten minute ThunderKAT observations since MeerTRAP started observing commensally with ThunderKAT in September 2021. This means that there has been approximately 17 hours of MeerTRAP monitoring of the field. We have not detected any further bursts from \mtpfrb\ with MeerTRAP. We found two low S/N image-plane transient candidates in 8\ s images of the field prior to September 2021. Due to the low S/N of these sources and the 8\ s time resolution, we cannot confirm whether these are artefacts or repeat bursts from the FRB source. MeerTRAP will continue to observe commensally with ThunderKAT during their weekly observations of \gx.

\subsection{Future FRB localisation prospects}
\label{sec: future prospects}

\mtpfrb\ is the first sub-arcsecond localised FRB using the MeerKAT telescope, and this localisation was made possible by the commensal nature of the MeerTRAP project. As MeerTRAP observes commensally with all MeerKAT LSPs, we can use the 8s images from the LSP observations to localise bright FRBs. This demonstrates the power of commensal searches for single pulses. However, only FRBs with high S/N can be localised in second-long images. We are currently working to implement a fast imaging pipeline as part of MeerTRAP. When MeerTRAP is triggered by a single pulse we can save the transient buffer data as phase delay corrected voltages. These data are typically $<1$s long, and we slice around the DM of the burst. We can then image these data offline to rapidly produce images and localise the burst. This pipeline will be an important tool for localising MeerTRAP FRBs and expanding the sample of FRBs associated with host galaxies.


\section{Conclusions}
\label{sec: conclusions}

We detected a new FRB, {using MeerKAT and the MeerTRAP backend. At the time of \mtpfrb's detection,} MeerTRAP was observing commensally with ThunderKAT, which enabled us to use the ThunderKAT 8s images to localise the FRB with sub-arcsecond precision. 
We used the specific luminosity and chance coincidence probability to determine that \mtpfrb\ is most likely of extra-galactic origin and is associated with nearby galaxy \irgal. We determined that the host galaxy is a disk/spiral galaxy at a redshift of $z = 0.066$. {However, there is a discrepancy between the \textsc{ymw16} and \textsc{ne2001} predictions for the Milky Way DM which means that there could be some uncertainty about its true location based on the measure DM alone. However the combination of the association with the nearby host galaxy, the predicted luminosity based on it being located in the proposed host, and the scattering properties support it as being extra-galactic.  This FRB highlights the uncertainty in the Milky Way DM and halo models that should be further explored with a larger sample of localised FRBs.} \mtpfrb\ is the first FRB localised to sub-arcsecond precision using MeerKAT.



\section*{Acknowledgements}

We thank Chris Williams for all his assistance in getting the MeerTRAP pipeline up and running and Jim Cordes for helpful discussions. We would like to thank the operators, SARAO staff and ThunderKAT Large Survey Project team.
LND, MC, FJ, KMR, BWS, MCB, MM, VM, SS, and MPS acknowledge support from the European Research Council (ERC) under the European Union's Horizon 2020 research and innovation programme (grant agreement No 694745).
MC acknowledges support of an Australian Research Council Discovery Early Career Research Award (project number DE220100819) funded by the Australian Government and the Australian Research Council Centre of Excellence for All Sky Astrophysics in 3 Dimensions (ASTRO 3D), through project number CE170100013.
RKK and HC acknowledge the support the South African Department of Science and Innovation and the National Research Foundation received through the SARAO SARCHI Research Chair.
HC is supported by Key Research Project of Zhejiang Lab (No. 2021PE0AC03).
DAHB acknowledges support from the South African National Research Foundation.
PAW acknowledges financial support from the NRF (grant no: 129359) and UCT.
MG is supported by the EU Horizon 2020 research and innovation programme under grant agreement No 101004719.
The MeerKAT telescope is operated by the South African Radio Astronomy Observatory (SARAO), which is a facility of the National Research Foundation, an agency of the Department of Science and Innovation.
We acknowledge use of the Inter-University Institute for Data Intensive Astronomy (IDIA) data intensive research cloud for data processing. IDIA is a South African university partnership involving the University of Cape Town, the University of Pretoria and the University of the Western Cape.
 The authors also acknowledge the usage of TRAPUM infrastructure funded and installed by the Max-Planck-Institut für Radioastronomie and the Max-Planck-Gesellschaft.
The SALT observations were obtained under the SALT Large Science Programme on transients (2018-2-LSP-001; PI: DAHB). Polish participation in SALT is funded by grant No. MEiN nr 2021/WK/01. 
This work has made use of data from the European Space Agency (ESA) mission
{\it Gaia}.\footnote{\href{https://www.cosmos.esa.int/gaia}{https://www.cosmos.esa.int/gaia}, processed by the {\it Gaia}
Data Processing and Analysis Consortium (DPAC,
\href{https://www.cosmos.esa.int/web/gaia/dpac/consortium}{https://www.cosmos.esa.int/web/gaia/dpac/consortium})} Funding for the DPAC
has been provided by national institutions, in particular the institutions
participating in the {\it Gaia} Multilateral Agreement.
This research made use of Astropy,\footnote{\href{http://www.astropy.org}{http://www.astropy.org}} a community-developed core Python package for Astronomy \citep{astropy:2013, astropy:2018}.
This research made use of APLpy, an open-source plotting package for Python \citep{2012ascl.soft08017R}.
This research uses services or data provided by the Astro Data Lab at NSF's National Optical-Infrared Astronomy Research Laboratory. NOIRLab is operated by the Association of Universities for Research in Astronomy (AURA), Inc. under a cooperative agreement with the National Science Foundation.
This research has made use of the VizieR catalogue access tool, CDS, Strasbourg, France\footnote{\href{10.26093/cds/vizier}{10.26093/cds/vizier}}. The original description 
of the VizieR service was published in \citet{2000A&AS..143...23O}.
This research has made use of the SIMBAD database,
operated at CDS, Strasbourg, France \citep{2000A&AS..143....9W}.
This research has made use of NASA’s Astrophysics Data System Bibliographic Services\footnote{\href{https://ui.adsabs.harvard.edu/}{https://ui.adsabs.harvard.edu/}}.
LND would like to acknowledge the traditional owners of the land where most of her work was performed: the Wurundjeri People of the Woi worrung Nation, the Whadjuk people of the Noongar Nation and the Gadigal People of the Eora Nation.
We would like to thank the referee for their constructive comments and help in improving this manuscript.


\section*{Data availability}

AstroAccelerate is available here: \href{https://github.com/AstroAccelerateOrg/astro-accelerate}{https://github.com/AstroAccelerateOrg/astro-accelerate}.
The \texttt{LOFAR Transients Pipeline} is available here: \href{https://tkp.readthedocs.io/en/latest/index.html}{https://tkp.readthedocs.io/en/latest/index.html}.
The code for performing the astrometry is available on Zenodo: \href{https://doi.org/10.5281/zenodo.4921715}{https://doi.org/10.5281/zenodo.4921715}.



\bibliographystyle{mnras}
\bibliography{mtp19} 



\appendix

\section{Repeat observations of the \gx\ field}
\label{appendix}

\begin{table}
    \centering
    \caption{Date of each ThunderKAT observation of \gx\ where we have imaged the epoch in 8\ s intervals. The number of 8\ s intervals that have been imaged per epoch is shown here.}
    \label{tab: ThunderKAT 8 second images}
    \begin{tabular}{lp{0.6cm}lp{0.6cm}lp{0.6cm}}
Date & \# of images & Date & \# of images & Date & \# of images \\
\hline
\hline
2018.09.28 & 27 & 2019.05.25 & 75 & 2020.01.03 & 75 \\
2018.10.05 & 109 & 2019.05.31 & 74 & 2020.01.10 & 73 \\
2018.10.11 & 112 & 2019.06.08 & 73 & 2020.01.20 & 75 \\
2018.10.12 & 112 & 2019.06.16 & 74 & 2020.01.25 & 74 \\
2018.10.19 & 109 & 2019.06.24 & 74 & 2020.02.02 & 73 \\
2018.10.27 & 111 & 2019.06.30 & 75 & 2020.02.08 & 74 \\
2018.11.03 & 112 & 2019.07.07 & 74 & 2020.02.15 & 75 \\
2018.11.10 & 110 & 2019.07.14 & 73 & 2020.02.21 & 74 \\
2018.11.17 & 112 & 2019.07.22 & 75 & 2020.03.02 & 73 \\
2018.11.24 & 112 & 2019.07.27 & 74 & 2020.03.09 & 74 \\
2018.12.08 & 111 & 2019.08.04 & 74 & 2020.03.14 & 75 \\
2018.12.15 & 112 & 2019.08.10 & 75 & 2020.03.21 & 74 \\
2018.12.22 & 112 & 2019.08.16 & 74 & 2020.03.29 & 74 \\
2018.12.29 & 111 & 2019.08.23 & 56 & 2020.04.03 & 31 \\
2019.01.05 & 112 & 2019.08.31 & 75 & 2020.04.10 & 75 \\
2019.01.12 & 112 & 2019.09.07 & 74 & 2020.04.18 & 73 \\
2019.01.19 & 112 & 2019.09.14 & 74 & 2020.04.25 & 74 \\
2019.01.26 & 112 & 2019.09.21 & 74 & 2020.05.01 & 74 \\
2019.02.01 & 111 & 2019.09.29 & 74 & 2020.05.09 & 74 \\
2019.02.09 & 111 & 2019.10.06 & 74 & 2020.05.17 & 74 \\
2019.02.16 & 111 & 2019.10.12 & 74 & 2020.06.08 & 75 \\
2019.02.23 & 111 & 2019.10.19 & 74 & 2020.06.14 & 74 \\
2019.03.01 & 11 & 2019.10.26 & 74 & 2020.06.17 & 73 \\
2019.03.09 & 111 & 2019.11.01 & 73 & 2020.06.19 & 74 \\
2019.03.18 & 72 & 2019.11.10 & 74 & 2020.06.26 & 74 \\
2019.03.25 & 74 & 2019.11.17 & 74 & 2020.07.04 & 74 \\
2019.04.01 & 59 & 2019.11.24 & 74 & 2020.07.12 & 75 \\
2019.04.09 & 64 & 2019.11.30 & 74 & 2020.07.19 & 74 \\
2019.04.15 & 75 & 2019.12.02 & 111 & 2020.07.26 & 75 \\
2019.04.20 & 74 & 2019.12.07 & 74 & 2020.08.01 & 74 \\
2019.04.29 & 73 & 2019.12.15 & 74 & 2020.08.08 & 74 \\
2019.05.05 & 60 & 2019.12.20 & 74 & 2020.08.15 & 74 \\
2019.05.11 & 72 & 2019.12.28 & 74 & 2020.08.22 & 75 \\
2019.05.18 & 74 &            &    &            &    \\
\hline
    \end{tabular}
\end{table}


\bsp	
\label{lastpage}
\end{document}